\documentclass[a4paper,11pt]{article}
\usepackage{pos}
\usepackage{tikz-cd}
\usepackage[colorinlistoftodos]{todonotes}

\newcommand{\RR}{\mathbb{R}}

\def\cross{\mbox{$\rule{0.7pt}{1.3ex}\!\times $}}

\title{Twisted geometry for submanifolds of $\mathbb{R}^n$}

\author*[a,b]{Gaetano Fiore}
\author[c,d]{Thomas Weber}

\affiliation[a]{Dip. di Matematica e Applicazioni, Universitá di Napoli “Federico II”, Complesso Universitario
MSA, Via Cintia, 80126 Naples, Italy}

\affiliation[b]{I.N.F.N., Sezione di Napoli, Complesso Universitario MSA, Via Cintia, 80126 Naples, Italy}

\affiliation[c]{Dipartimento di Scienze e Innovazione Tecnologica, Universit\'a degli Studi del Piemonte Orientale
“Amedeo Avogadro”, Viale Teresa Michel 11, 15121 Alessandria, Italy}

\affiliation[d]{I.N.F.N., Sezione di Torino, Via P. Giuria 1, 10125 Turin, Italy}

\emailAdd{gaetano.fiore@unina.it}
\emailAdd{thomas.weber@uniupo.it}

\abstract{This is a friendly introduction to our recent general procedure for constructing noncommutative deformations
of an embedded submanifold $M$ of $\RR^n$ determined by a set of smooth  equations $f^a(x)=0$. We use the framework of Drinfel'd  twist deformation of differential geometry pioneered in [Aschieri et al.,
Class. Quantum Gravity 23 (2006), 1883]; the commutative pointwise product
is replaced by a (generally noncommutative) $\star$-product induced by a  Drinfel'd twist.}

\FullConference{%
  Corfu Summer Institute 2021 "School and Workshops on Elementary Particle Physics and Gravity"\\
  29 August - 9 October 2021\\
  Corfu, Greece
}

\tableofcontents

\begin{document}
\maketitle

\section{Introduction}

Nowadays noncommutative Geometry  (NCG) \cite{Connes,GraFigVar00,Woronowicz1989,Lan97,Madore99,Majid2000}
is a broad research field aiming, among other things, at formulating candidate frameworks for the
quantization of gravity  (see e.g. \cite{DopFreRob95,Aschieri2006}) or
the unification of fundamental interactions (see e.g. \cite{ConLot91,ChaConvan,AscMadManSteZou}). 
It is natural to ask whether and to what extent
the  notion of a submanifold,
which is ubiquitous in mathematics and physics (think e.g. of:
equipotential hypersurfaces; wavefronts for wave equations; submanifolds where to impose initial or boundary conditions for fields defined on the encompassing manifold; ADS/CFT  correspondence and the holographic principle;
 lightcones, event horizons and other  null hypersurfaces in general relativity, etc.)
can be generalized from classical differential geometry to NCG. So far these questions have been answered by making sense of many special examples of noncommutative (NC) submanifolds\footnote{For instance, the noncommutative algebra $\mathcal{A}$ ``of functions on the quantum group $SU_q(n)$" is obtained from 
the one  on the quantum group $U_q(n)$  by imposing that the so-called $q$-determinant be 1, as in the $q=1$ commutative  limit, 
and one can construct various differential calculi on $\mathcal{A}$ \cite{Woronowicz1989}.},
but have not received sufficient general treatment, except in few articles (see e.g.
\cite{Masson1995,Giunashvili,TWeber2019,FioFraWebquadrics,FioreWeber}).
This proceeding summarizes the contributions to the topic of Ref. \cite{FioFraWebquadrics,FioreWeber}, which address the above questions systematically within the framework
of deformation quantization \cite{BayFlaFroLicSte}, in the particular approach based on  Drinfel'd twisting \cite{Drinfeld1983} of Hopf algebras, for  embedded submanifolds $M$ of $\mathbb{R}^n$  consisting  of points of $x$  fulfilling a set of equations  
\begin{equation}
f^a(x)=0,\qquad a=1,2,...,k<n.                 \label{DefIdeal}
\end{equation}
Here $f\equiv(f^1,...,f^k):\RR^n\rightarrow\RR^k$ are smooth 
functions such that the Jacobian matrix $J=\partial f/\partial x$ is of rank $k$ 
on all $\RR^n$; or, more generally, where $f$ is well-defined and $J$ is of rank $k$ on an open
subset ${\cal D}_f\subset\RR^n$,  and $M$
consists  of the points of  ${\cal D}_f$  fulfilling (\ref{DefIdeal}).
In fact, in \cite{FioFraWebquadrics,FioreWeber} one obtains NC deformations of the geometry on a whole {\it $k$-parameter family} of embedded submanifolds $M_c:=f^{-1}(c)\subset {\cal D}_f$ [with $c \equiv(c^1,...,c^k)\in f\left({\cal D}_f\right)$, \ $M_0=M$] of dimension $n\!-\!k$; each $M_c$ is the level set of $f$ consisting of points $x$ such that $ f^a_c(x):=f^a(x)-c^a=0$
for all $a=1,\ldots,k$. 
Embedded submanifolds $N\subset M$ can be obtained by
adding more equations to (\ref{DefIdeal}).

In deformation quantization \cite{BayFlaFroLicSte} the commutative algebra
$\mathcal{A}=\mathcal{C}^\infty(\mathbb{R}^n)$ of smooth functions on a smooth manifold
$\mathbb{R}^n$ is replaced by a star product algebra $\mathcal{A}_\star=
(\mathcal{C}^\infty(\mathbb{R}^n)[[\nu]],\star)$, modelled on the formal power series 
$\mathcal{C}^\infty(\mathbb{R}^n)[[\nu]]$ in a deformation parameter $\nu$.  $\star$  deforms the pointwise
product $m(f\otimes g)=fg$ of functions $f,g\in\mathcal{A}$, 
$f\star g=fg+\mathcal{O}(\nu)$, while staying 
associative and unital.
In the case of Drinfel'd twist deformation quantization \cite{Aschieri2006,Drinfeld1983}
any  normalized $2$-cocycle
\begin{equation}
\label{twist}
    \mathcal{F}=1\otimes 1+\mathcal{O}(\nu)\in(U\Xi\otimes U\Xi)[[\nu]]
\end{equation}
(a {\it twist}) on the enveloping algebra $U\Xi$ of the Lie algebra $\Xi$ of vector fields
(identified with first order differential operators)  on $\mathbb{R}^n$
induces a twist star product $\star:=m\circ\mathcal{F}^{-1}(\rhd\otimes\rhd)$ on $\mathbb{R}^n$,
where $\rhd$ is the extension of the Lie derivative. 
This process is functorial  \cite{HenSte2000}, i.e. $\mathcal{F}$ deforms 
$\mathcal{A}=\mathcal{C}^\infty(\mathbb{R}^n)$-modules into $\mathcal{A}_\star$-modules,
and  $\mathcal{A}$-linear operations into $\mathcal{A}_\star$-linear operations.
In particular, the $\mathcal{A}$-bimodules of vector fields $\Xi$ and differential forms $\Omega$ on $\mathbb{R}^n$ are
deformed into $\mathcal{A}_\star$-bimodules.  $\star$-Lie derivatives
are twisted derivations and one  obtains a twisted Cartan calculus \cite{Aschieri2006}.
The guiding idea of the notion of NC submanifolds in this setting is best
explained by the commutativity of the  diagram
\begin{equation}\label{Diag}
\begin{tikzcd}
\mathcal{A}=\mathcal{C}^\infty(\mathbb{R}^n)
\arrow{rrrr}{\text{Submanifold Projection}}
\arrow{d}[swap]{\text{Quantization}}
& & & &\mathcal{B}=\mathcal{C}^\infty(M)
\arrow{d}{\text{Quantization}}\\
\mathcal{A}_\star=(\mathcal{C}^\infty(\mathbb{R}^n)[[\nu]],\star)
\arrow{rrrr}{\text{Submanifold Projection}}
& & & &\mathcal{B}_\star=(\mathcal{C}^\infty(M)[[\nu]],\star')
\end{tikzcd}
\end{equation}
In words, we induce a quantization of a submanifold $M$
via a quantization of the manifold $\mathbb{R}^n$, given the commutativity of (\ref{Diag}).
As said, in \cite{FioFraWebquadrics,FioreWeber} we are interested in
the situation when
$M$ is a submanifold given in terms of generators $(x^1,\ldots,x^n)$ and relations
(\ref{DefIdeal}).
We show that, in case the deformation $\mathcal{A}_\star$ is obtained by
a twist $\mathcal{F}$ based on the Lie algebra $\Xi_t$ of vector fields tangent to {\it all} the $M_c$,
the twist star product on $M$ makes the diagram (\ref{Diag}) commute. If $\mathcal{F}$
is even based on vector fields in $\Xi_t$ that are Killing\footnote{This restriction might be relaxed
by adopting the more general framework recently introduced in \cite{Aschieri2020}.} 
for a given (pseudo)Riemannian metric on $\mathbb{R}^n$,
the twist deformation extends to the level of (pseudo)Riemannian geometry so that
quantization and submanifold projection commute.
Furthermore, in the case of quadrics $M$ embedded in $\RR^n$,
we give explicit descriptions of both star product algebras  $\mathcal{A}_\star$, $\mathcal{B}_\star$,
as well as of the corresponding twisted vector fields and differential forms, via twisted generators and
relations. Examples of  codimension $2$ twisted submanifold will appear in \cite{GTCodimension2}. 
Note that the presented procedure is a \textit{global} approach,
i.e. we consider the algebra of global functions or bimodules of global sections of a bundle and deform them as such. One way to take locality into account is
given by the sheaf-theoretic approach to NC calculi on subalgebras proposed in
\cite{AschieriFioresiLatiniWeber2021}.

The proceeding is organized as follows. In Chapter~\ref{Chap2} we  recall the notions of
Hopf $*$-algebras and their representations (Section \ref{HopfA}), of their  twist deformations (Section \ref{DrinfeldTwists}), 
of twisted Cartan calculus (Section \ref{TwistedCC}) and Riemannian geometry (Section \ref{sectRG}).
The first part of Chapter~\ref{Chap3} concerns the twist deformation of submanifolds of $\mathbb{R}^n$,
as discussed above;  in  Section~\ref{TwistedQuad}
we present an explicit treatment of twisted quadrics of $\mathbb{R}^3$, focusing on the family of hyperboloids and cone, especially the circular ones in $\RR^3$ endowed with Minkowski metric.

\section{Twisted Riemannian geometry}\label{Chap2}

\subsection{Hopf \texorpdfstring{$*$}{*}-algebras and their representations}
\label{HopfA}

In the following $\mathbb{K}$ denotes the field or real numbers or the field of complex
numbers. Fix a \textit{Hopf $*$-algebra} $(H,\Delta,\epsilon,S,*)$ with coproduct
$\Delta\colon H\rightarrow H\otimes H$, counit $\epsilon\colon H\rightarrow\mathbb{K}$,
antipode $S\colon H\rightarrow H$ and $*$-involution $*\colon H\rightarrow H$.
The latter is an antilinear, involutive, anti-algebra map satisfying
\begin{equation}
    (*\otimes *)\circ\Delta=\Delta\circ*,\hspace{1cm}
    \epsilon\circ *={}^{\overline{~}}\circ\epsilon,\hspace{1cm}
    S\circ*\circ S\circ*=\mathrm{id}_H,
\end{equation}
where ${}^{\overline{~}}\colon H\rightarrow H$ denotes the complex conjugation.
The main class of examples we are interested in is that of the universal enveloping
algebra $U\mathfrak{g}$ of a \textit{$*$-Lie algebra} $\mathfrak{g}$. Here 
$(\mathfrak{g},[\cdot,\cdot])$ is a Lie algebra together with an antilinear, involutive
map $*\colon\mathfrak{g}\rightarrow\mathfrak{g}$ such that $[x,y]^*=[y^*,x^*]$ for all
$x,y\in\mathfrak{g}$. After extension as an anti-algebra homomorphism $*$ constitutes a
$*$-involution on $U\mathfrak{g}$, compatible with the usual coproduct, counit and antipode
on $U\mathfrak{g}$, which are determined on primitive elements $x\in\mathfrak{g}$ via
\begin{equation}
    \Delta(x)=x\otimes 1+1\otimes x,\hspace{1cm}
    \epsilon(x)=0,\hspace{1cm}
    S(x)=-x.
\end{equation}
The representation theory of a Hopf $*$-algebra concerns \textit{$H$-$*$-modules}, 
namely left $H$-modules $(\mathcal{M},\rhd)$
together with a $*$-involution on $\mathcal{M}$,
denoted by the same symbol for simplicity, such that
\begin{equation}
    (h\rhd s)^*=S(h)^*\rhd s^*
\end{equation}
for all $h\in H$ and $s\in\mathcal{M}$. Morphisms of left $H$-$*$-modules are
left $H$-module morphisms that intertwine the $*$-involutions.
A \textit{left $H$-module $*$-algebra} is a $*$-algebra $\mathcal{A}$ endowed with a left 
$H$-$*$-module structure $\rhd\colon H\otimes\mathcal{A}\rightarrow\mathcal{A}$
such that for all $a,b\in\mathcal{A}$ and $h\in H$
\begin{equation}
    h\rhd(ab)=(h_{(1)}\rhd a)(h_{(2)}\rhd b),\hspace{1cm}
    h\rhd 1_\mathcal{A}=\epsilon(h)1_\mathcal{A},
\end{equation}
where we utilize Sweedler's summation $\Delta(h)=:h_{(1)}\otimes h_{(2)}$.
For a left $H$-module $*$-algebra $(\mathcal{A},*,\rhd)$ we call an $\mathcal{A}$-bimodule
$\mathcal{M}$ an \textit{$H$-equivariant $\mathcal{A}$-$*$-bimodule} if $\mathcal{M}$
is a left $H$-$*$-module such that
\begin{equation}
    h\rhd(a\cdot s\cdot b)=(h_{(1)}\rhd a)\cdot(h_{(2)}\rhd s)
    \cdot(h_{(3)}\rhd b),\hspace{1cm}
    (a\cdot s\cdot b)^*=b^*\cdot s^*\cdot a^*
\end{equation}
for all $h\in H$, $a,b\in\mathcal{A}$ and $s\in\mathcal{M}$.
By a slight abuse of notation we denoted the left $H$-module action and $*$-involution on
$\mathcal{M}$ the same way as for $\mathcal{A}$, while we used $\cdot$ for the
left and right module action of $\mathcal{A}$ on $\mathcal{M}$.
The notions of left $H$-$*$-module, left $H$-module $*$-algebra
and $H$-equivariant $\mathcal{A}$-$*$-bimodule extend to
$\mathbb{N}_0$-graded vector spaces by demanding the 
corresponding actions and $*$-involutions to be graded maps.

If $H$ is \textit{cocommutative}, i.e. $\Delta^\mathrm{op}=\Delta$, the category of left $H$-$*$-modules admits a symmetric monoidal structure,
where we endow the tensor product
$\mathcal{M}\otimes\mathcal{N}$ of two left $H$-$*$-modules $\mathcal{M},\mathcal{N}$
with the left $H$-action and $*$-involution
\begin{equation}\label{eq01}
    h\rhd(s\otimes t):=(h_{(1)}\rhd s)\otimes(h_{(2)}\rhd t),\hspace{1cm}
    (s\otimes t)^*:=s^*\otimes t^*
\end{equation}
defined for all $h\in H$, $s\in\mathcal{M}$ and $t\in\mathcal{N}$.
The isomorphism $\tau\colon\mathcal{M}\otimes \mathcal{N}\to \mathcal{N}\otimes \mathcal{M}$ defined by $\tau(s\otimes t)=t\otimes s$,
is the corresponding symmetric braiding.

In case $\mathcal{A}$ is a commutative left $H$-module $*$-algebra for $H$ cocommutative
we structure the category of symmetric $H$-equivariant
$\mathcal{A}$-$*$-bimodules
as a symmetric monoidal category using the tensor product $\otimes_\mathcal{A}$.
Let $\mathcal{M},\mathcal{N}$ be such symmetric $H$-equivariant
$\mathcal{A}$-$*$-bimodules. Here, symmetric means that 
$a\cdot s=s\cdot a$ for all $a\in \mathcal{A}$ and $s\in\mathcal{M}$. The left $H$-module action
and $*$-involution on $\mathcal{M}\otimes_\mathcal{A}\mathcal{N}$ are induced from
(\ref{eq01}) and again the braiding is given by the tensor flip. 

For any $\mathbb{K}$-vector space $V$ the formal power series $V[[\nu]]$ in a formal
parameter $\nu$ are a $\mathbb{K}[[\nu]]$-module and we can extend any $\mathbb{K}$-linear
map $f\colon V\rightarrow W$ to a $\mathbb{K}[[\nu]]$-linear map
$V[[\nu]]\rightarrow W[[\nu]]$, denoted by the same symbol $f$. As a consequence
any Hopf $*$-algebra $H$ over $\mathbb{K}$ can be extended to a 
Hopf $*$-algebra $H[[\nu]]$ over $\mathbb{K}[[\nu]]$, where we have to employ the
completed tensor product in the $\nu$-adic topology.

\subsection{Drinfel'd twists and twisted representations}
\label{DrinfeldTwists}

Fix a Hopf $*$-algebra $H$. A \textit{(Drinfel'd) twist} on $H$ is an element
$\mathcal{F}=1\otimes 1+\mathcal{O}(\nu)\in(H\otimes H)[[\nu]]$ satisfying
$2$-cocycle and normalization condition
\begin{equation}
\begin{split}
    (\mathcal{F}\otimes 1)(\Delta\otimes\mathrm{id})(\mathcal{F})
    =&(1\otimes\mathcal{F})(\mathrm{id}\otimes\Delta)(\mathcal{F}),\\
    (\epsilon\otimes\mathrm{id})(\mathcal{F})
    =&1=(\mathrm{id}\otimes\epsilon)(\mathcal{F}).
\end{split}
\end{equation}
We frequently use \textit{leg notation} $\mathcal{F}=\mathcal{F}_1\otimes\mathcal{F}_2$
and similarly $\overline{\mathcal{F}}=
\overline{\mathcal{F}}_1\otimes\overline{\mathcal{F}}_2$ for the inverse
$\overline{\mathcal{F}}$ of $\mathcal{F}$. If several copies of $\mathcal{F}$ or its
inverse appear we write $\overline{\mathcal{F}}=\overline{\mathcal{F}}_{1'}\otimes
\overline{\mathcal{F}}_{2'}$ for the second copy, et cetera, to distinguish the different
summations. For any twist we define $\beta:=\mathcal{F}_1S(\mathcal{F}_2)\in H[[\nu]]$
and $\beta^{-1}:=S(\overline{\mathcal{F}}_1)\overline{\mathcal{F}}_2\in H[[\nu]]$.
One can show that $\beta^{-1}$ is in fact the inverse of $\beta$. A twist
$\mathcal{F}$ is said to be 
\begin{enumerate}
\item[$\bullet$] \textit{real} if $\mathcal{F}_1^*\otimes\mathcal{F}_2^*
=S(\mathcal{F}_2)\otimes S(\mathcal{F}_1)$ \cite{Aschieri2006}  and

\item[$\bullet$] \textit{unitary} if $\mathcal{F}_1^*\otimes\mathcal{F}_2^*
=\overline{\mathcal{F}}_1\otimes\overline{\mathcal{F}}_2$ \cite{Fiore2010} .
\end{enumerate}
Assume that the Hopf $*$-algebra $H$ is cocommutative. Consider a commutative
left $H$-module $*$-algebra $\mathcal{A}$ and a 
symmetric $H$-equivariant $\mathcal{A}$-$*$-bimodule $\mathcal{M}$. In the following
we deform the given data using a real or unitary twist $\mathcal{F}$ on $H$.
First we construct the \textit{twisted Hopf algebra} $H^\mathcal{F}$ as the algebra
$H[[\nu]]$ with extended counit, but coproduct and antipode given by
\begin{equation}
    \Delta_\mathcal{F}(h):=\mathcal{F}\Delta(h)\overline{\mathcal{F}}\hspace{1cm}
    \text{and}\hspace{1cm}
    S_\mathcal{F}(h):=\beta S(h)\beta^{-1}
\end{equation}
for all $h\in H$. If $\mathcal{F}$ is real the Hopf algebra $H^\mathcal{F}$ becomes a
Hopf $*$-algebra with respect to the $*$-involution 
$h^{*_\mathcal{F}}=\beta h^*\beta^{-1}$ for all $h\in H^\mathcal{F}$. For a unitary
twist $H^\mathcal{F}$ is a Hopf $*$-algebra with respect to the undeformed $*$-involution.

The twist deformation $\mathcal{A}_\star$ of $\mathcal{A}$ is the 
$\mathbb{K}[[\nu]]$-module $\mathcal{A}[[\nu]]$ endowed with the same unit and deformed
product $a\star b:=(\overline{\mathcal{F}}_1\rhd a)(\overline{\mathcal{F}}_2\rhd b)$
for all $a,b\in\mathcal{A}_\star$. It is a left $H^\mathcal{F}$-module algebra, i.e.
\begin{equation}
    h\rhd(a\star b)=(h_{\widehat{(1)}}\rhd a)\star(h_{\widehat{(2)}}\rhd b),
\end{equation}
where we denoted the twisted product by $\Delta_\mathcal{F}(h)=:
h_{\widehat{(1)}}\otimes h_{\widehat{(2)}}$. In addition, $\mathcal{A}_\star$ is
\textit{twisted commutative}, i.e. $b\star a
=(\overline{\mathcal{R}}_1\rhd a)\star(\overline{\mathcal{R}}_2\rhd b)$, where
$\overline{\mathcal{R}}=\overline{\mathcal{R}}_1\otimes\overline{\mathcal{R}}_2$
is the inverse of the \textit{braiding}
$\mathcal{R}:=\mathcal{R}_1\otimes\mathcal{R}_2:=\mathcal{F}_2\overline{\mathcal{F}}_{1'}
\otimes\mathcal{F}_1\overline{\mathcal{F}}_{2'}\in H^\mathcal{F}\otimes H^\mathcal{F}$.
If $\mathcal{A}$ is $\mathbb{N}_0$-graded and graded-commutative,
i.e. $ab=(-1)^{|a|\cdot|b|}ba$, where $|a|,|b|$ denote the degrees of $a,b\in\mathcal{A}$, then $\mathcal{A}_\star$ is \textit{twisted graded-commutative}, i.e. $b\star a
=(-1)^{|a|\cdot|b|}(\overline{\mathcal{R}}_1\rhd a)\star(\overline{\mathcal{R}}_2\rhd b)$ for all $a,b\in\mathcal{A}_\star$.
The braiding $\mathcal{R}$ satisfies $\mathcal{R}_2\otimes\mathcal{R}_1=\overline{\mathcal{R}}$ 
and
the \textit{hexagon equations}
\begin{equation}
    (\Delta_\mathcal{F}\otimes\mathrm{id})(\mathcal{R})=\mathcal{R}_1\otimes\mathcal{R}_{1'}
    \otimes\mathcal{R}_2\mathcal{R}_{2'}
    \hspace{1cm}\text{and}\hspace{1cm}
    (\mathrm{id}\otimes\Delta_\mathcal{F})(\mathcal{R})=\mathcal{R}_1\mathcal{R}_{1'}
    \otimes\mathcal{R}_{2'}\otimes\mathcal{R}_2.
\end{equation}
For real twists $\mathcal{R}_1^{*_\mathcal{F}}\otimes\mathcal{R}_2^{*_\mathcal{F}}
=\overline{\mathcal{R}}$, while for unitary twists $\mathcal{R}_1^*\otimes\mathcal{R}_2^*
=\overline{\mathcal{R}}$. If $\mathcal{F}$ is real then 
$\mathcal{A}_\star$ is a left $H^\mathcal{F}$-module $*$-algebra with respect to the
undeformed $*$-involution (while we have to twist the $*$-involution of $H^\mathcal{F}$).
On the other hand, if $\mathcal{F}$ is unitary (i.e. the $*$-involution of 
$H^\mathcal{F}$ is undeformed) $\mathcal{A}_\star$ becomes a
left $H^\mathcal{F}$-module 
$*$-algebra via $a^{*_\star}:=S(\beta)\rhd a^*$ for all $a\in\mathcal{A}_\star$.

Similarly, the twist deformation $\mathcal{M}_\star$ of $\mathcal{M}$ is described as
the $\mathbb{K}[[\nu]]$-module $\mathcal{M}[[\nu]]$ together with the $\mathcal{A}_\star$-module actions
\begin{equation}
    a\star s:=(\overline{\mathcal{F}}_1\rhd a)\cdot(\overline{\mathcal{F}}_2\rhd s)
    \hspace{1cm}\text{and}\hspace{1cm}
    s\star a:=(\overline{\mathcal{F}}_1\rhd s)(\overline{\mathcal{F}}_2\rhd a)
\end{equation}
for all $a\in\mathcal{A}_\star$ and $s\in\mathcal{M}[[\nu]]$. Together with the
$\mathbb{K}[[\nu]]$-linearly extended left $H^\mathcal{F}$-action $\mathcal{M}_\star$
is an $H_\star$-equivariant $\mathcal{A}_\star$-bimodule. Furthermore, it is
\textit{twisted symmetric}, i.e.
\begin{equation}\label{eq11}
    s\star a=(\overline{\mathcal{R}}_1\rhd a)\star(\overline{\mathcal{R}}_2\rhd s)
\end{equation}
for all $s\in\mathcal{M}_\star$ and $a\in\mathcal{A}_\star$.
If $\mathcal{F}$ is real then 
$\mathcal{M}_\star$ is an $H^\mathcal{F}$-equivariant $\mathcal{A}_\star$-$*$-bimodule
with respect to the
undeformed $*$-involution, while
if $\mathcal{F}$ is unitary $\mathcal{M}_\star$ becomes an 
$H^\mathcal{F}$-equivariant $\mathcal{A}_\star$-$*$-bimodule 
via $s^{*_\star}:=S(\beta)\rhd s^*$ for all $s\in\mathcal{M}_\star$.

For two left $H$-modules $\mathcal{M},\mathcal{N}$ the \textit{twisted tensor product}
$\mathcal{M}_\star\otimes_\star\mathcal{N}_\star$ is given by $(\mathcal{M}\otimes\mathcal{N})[[\nu]]$, where
$s\otimes_\star t:=(\overline{\mathcal{F}}_1\rhd s)
\otimes(\overline{\mathcal{F}}_2\rhd t)$ for all $s\in\mathcal{M}$ and $t\in\mathcal{N}$.
It follows that $\mathcal{M}_\star\otimes_\star\mathcal{N}_\star$ is a left $H^\mathcal{F}$-module
and one can show that the left $H^\mathcal{F}$-module isomorphism
\begin{equation}
    \sigma_{\mathcal{M},\mathcal{N}}\colon\mathcal{M}_\star\otimes_\star\mathcal{N}_\star\ni(s\otimes_\star t)\mapsto
    (\overline{\mathcal{R}}_1\rhd t)\otimes_\star(\overline{\mathcal{R}}_2\rhd s)\in \mathcal{N}_\star\otimes_\star\mathcal{M}_\star
\end{equation}
determines a symmetric braiding on the monoidal category of twisted left $H$-modules.
If $\mathcal{F}$ is real (respectively unitary) we structure
$\mathcal{M}_\star\otimes_\star\mathcal{N}_\star$ as a
left $H^\mathcal{F}$-$*$-module via
\begin{equation}
    (s\otimes_\star t)^*=(\overline{\mathcal{R}}_1\rhd s^*)\otimes_\star
    (\overline{\mathcal{R}}_2\rhd t^*),
    \text{ respectively }
    (s\otimes_\star t)^{*_\star}=(\overline{\mathcal{R}}_1\rhd s^{*_\star})\otimes_\star
    (\overline{\mathcal{R}}_2\rhd t^{*_\star}).
\end{equation}
Similar results hold for symmetric $H$-equivariant $\mathcal{A}$-bimodules $\mathcal{M},\mathcal{N}$, using
$\mathcal{M}_\star\otimes_{{\cal A}_\star}\mathcal{N}_\star$.

\noindent
One can complete \cite{GurevichMajid1994,Aschieri2006}  the $H^\mathcal{F}$-module algebra $(H[[\nu]],\star)$ itself into a triangular Hopf algebra $H_\star\!=\!(H[[\nu]],\star,
\Delta_\star,\epsilon,S_\star,{\cal R}_\star)$  isomorphic  to $H^\mathcal{F}\!=\!(H[[\nu]],\cdot,
\Delta_\mathcal{F},\epsilon,S_\mathcal{F},{\cal R})$  (cf. also \cite{Fio98JMP,Fio00RMP,Fiore2010}).

Examples of unitary twists on $U\mathfrak{g}$ for a $*$-Lie algebra $\mathfrak{g}$ are
\begin{enumerate}
\item[$\bullet$] \textit{abelian twists} $\mathcal{F}=\exp(\mathrm{i}\nu P)$,
where $P=\frac{1}{2}\sum_i(e_i\otimes f_i-f_i\otimes e_i)$ \cite{Reshetikhin1990}
is a finite sum of pairwise commuting (anti-)Hermitian
elements $e_i,f_i\in\mathfrak{g}$ and

\item[$\bullet$] \textit{Jordanian twists} $\mathcal{F}=\exp\big(\frac{1}{2}H\otimes\log(1+\mathrm{i}\nu E)\big)$ \cite{Ogievetsky1992}, where
$H,E\in\mathfrak{g}$ are anti-Hermitian elements such that $[H,E]=2E$.
\end{enumerate}
The abelian twist is real, in addition.

\subsection{Twisted Cartan calculus}\label{TwistedCC}

Let us substantiate the previous twist deformation procedure via the concrete example
of the tensor algebra of a smooth manifold $M$. For the rest of the article we operate
in this framework. The algebra $\mathcal{X}:=\mathcal{C}^\infty(M)$ of smooth
$\mathbb{K}$-valued functions on $M$ is a commutative $*$-algebra with respect to
the pointwise product and the $*$-involution $f^*(p):=\overline{f(p)}$, 
where $f\in\mathcal{X}$ and $p\in M$, given by complex conjugation. Vector fields
$\Xi:=\Gamma^\infty(TM)$ on $M$ form a Lie $*$-algebra with respect to the
$*$-involution $\mathcal{L}_{X^*}f:=-(\mathcal{L}_Xf^*)^*$ for all 
$f\in\mathcal{X}$, where $\mathcal{L}_X$ denotes the Lie derivative of $X\in\Xi$.
This amplifies to the Hopf $*$-algebra $H:=U\Xi$, the latter acting on $\mathcal{X}$
via the Lie derivative, structuring $\mathcal{X}$
as a commutative left $H$-module $*$-algebra. More in general, the tensor algebra
$\mathcal{T}:=\bigoplus_{p,r\in\mathbb{N}_0}\mathcal{T}^{p,r}$, where
\begin{equation}
    \mathcal{T}^{p,r}:=
    \underbrace{\Omega\otimes_\mathcal{X}\ldots\otimes_\mathcal{X}\Omega}_{p\text{-times}}
    \otimes_\mathcal{X}
    \underbrace{\Xi\otimes_\mathcal{X}\ldots\otimes_\mathcal{X}\Xi}_{r\text{-times}}
\end{equation}
and $\Omega:=\Gamma^\infty(T^*M)$, is a symmetric $H$-equivariant 
$\mathcal{X}$-$*$-bimodule. The $\Xi$-action on $\mathcal{T}^{p,r}$ is obtained
by the Lie derivative
\begin{equation}\label{eq02}
\begin{split}
    X\rhd(\omega_1\otimes_\mathcal{X}&\ldots\otimes_\mathcal{X}\omega_p
    \otimes_\mathcal{X}Y_1\otimes_\mathcal{X}\ldots\otimes_\mathcal{X}Y_r)\\
    =&\mathcal{L}_{X_{(1)}}\omega_1\otimes_\mathcal{X}
    \ldots\otimes_\mathcal{X}\mathcal{L}_{X_{(p)}}\omega_p
    \otimes_\mathcal{X}\mathcal{L}_{X_{(p+1)}}Y_1\otimes_\mathcal{X}
    \ldots\otimes_\mathcal{X}\mathcal{L}_{X_{(p+r)}}Y_r,
\end{split}
\end{equation}
where $\omega_1,\ldots,\omega_p\in\Omega$, $X,Y_1,\ldots,Y_r\in\Xi$ and
$\mathcal{L}_X\omega_i=(\mathrm{i}_X\circ\mathrm{d}+\mathrm{d}\circ\mathrm{i}_X)\omega_i$,
$\mathcal{L}_XY_i=[X,Y_i]$. We extend (\ref{eq02}) as an $U\Xi$-action by
$\mathcal{L}_{XY}=\mathcal{L}_X\mathcal{L}_Y$ and $\mathcal{L}_{1_\mathbb{K}}=\mathrm{id}$
for all $X,Y\in\Xi$.

A unitary or real twist $\mathcal{F}$ on $H$ induces a twisted commutative
left $H^\mathcal{F}$-module $*$-algebra $\mathcal{X}_\star$ and a twisted symmetric
$H^\mathcal{F}$-equivariant $\mathcal{X}_\star$-$*$-bimodule $\mathcal{T}_\star$
according to the previous section. In more detail, $\mathcal{T}_\star
=\bigoplus_{p,r\in\mathbb{N}_0}\mathcal{T}_\star^{p,r}$ is defined by
\begin{equation}
    \mathcal{T}_\star^{p,r}:=
    \underbrace{\Omega_\star\otimes_{\mathcal{X}_\star}
    \ldots\otimes_{\mathcal{X}_\star}\Omega_\star}_{p\text{-times}}
    \otimes_{\mathcal{X}_\star}
    \underbrace{\Xi_\star\otimes_{\mathcal{X}_\star}
    \ldots\otimes_{\mathcal{X}_\star}\Xi_\star}_{r\text{-times}}
\end{equation}
and the $H^\mathcal{F}$-action is given by (\ref{eq02}), where we replace $\Delta$
by $\Delta_\mathcal{F}$.
Above, $\Omega_\star$ denotes the $\mathcal{X}_\star$-bimodule of twisted $1$-forms,
i.e. $\Omega_\star=\Omega[[\nu]]$ as $\mathbb{K}[[\nu]]$-modules and we endow the former
with the $\mathcal{X}_\star$-actions
$f\star\omega=(\overline{\mathcal{F}}_1\rhd f)\cdot(\overline{\mathcal{F}}_2\rhd\omega)$
and
$\omega\star f=(\overline{\mathcal{F}}_1\rhd\omega)\cdot(\overline{\mathcal{F}}_2\rhd f)$
for all $f\in\mathcal{X}_\star$ and $\omega\in\Omega_\star$. Similarly $\Xi$ is structured
as an $\mathcal{X}_\star$-bimodule and all the bimodules are twisted symmetric.
We understand the tensor product $\otimes_{\mathcal{X}_\star}$ with respect to this
$\mathcal{X}_\star$-bimodule structure, i.e. $(T_1\star f)\otimes_{\mathcal{X}_\star}T_2
=T_1\otimes_{\mathcal{X}_\star}(f\star T_2)$ for all $f\in\mathcal{X}_\star$ and
$T_1,T_2\in\mathcal{T}_\star$. The dual pairing
$\langle\cdot,\cdot\rangle\colon\Xi\otimes_\mathcal{X}\Omega\rightarrow\mathcal{X}$
deforms into an $\mathcal{X}_\star$-bilinear operation
\begin{equation}
    \langle\cdot,\cdot\rangle_\star
    :=\langle\cdot,\cdot\rangle\circ\overline{\mathcal{F}}\rhd
    \colon\Xi_\star\otimes_{\mathcal{X}_\star}\Omega_\star\rightarrow\mathcal{X}_\star.
\end{equation}
We choose $\star$-dual frames $\{e_i\}\subset\Xi_\star$ and
$\{\theta^i\}\subset\Omega_\star$, i.e. $\langle e_i,\theta^j\rangle_\star=\delta_i^j$
c.f. \cite{Aschieri2006}.
Employing the \textit{twisted Lie derivative} $\mathcal{L}^\mathcal{F}_\xi T
:=\mathcal{L}_{\overline{\mathcal{F}}_1\rhd\xi}(\overline{\mathcal{F}}_2\rhd T)$
for all $\xi\in H^\mathcal{F}$ and $T\in\mathcal{T}_\star$ we obtain a deformed action
of $H^\mathcal{F}$ on $\mathcal{T}_\star$. In particular, $\Xi_\star$ becomes a
\textit{twisted Lie algebra} via 
\begin{equation}
    [X,Y]_\star:=\mathcal{L}^\mathcal{F}_XY
    =[\overline{\mathcal{F}}_1\rhd X,\overline{\mathcal{F}}_2\rhd Y]
    =X\star Y-(\overline{\mathcal{R}}_1\rhd Y)\star(\overline{\mathcal{R}}_2\rhd X),
\end{equation}
i.e. $[Y,X]_\star=-[\overline{\mathcal{R}}_1\rhd X,\overline{\mathcal{R}}_2\rhd Y]_\star$
and $[X,[Y,Z]_\star]_\star=[[X,Y]_\star,Z]_\star+[\overline{\mathcal{R}}_1\rhd Y,
[\overline{\mathcal{R}}_2\rhd X,Z]_\star]_\star$
for all $X,Y,Z\in\Xi_\star$. The entirety of those structures is referred to as the
\textit{twisted Cartan calculus}, see \cite{Aschieri2006,TWeber2019} for more information.

\subsection{Twisted Riemannian geometry}\label{sectRG}

The process of twist deformation turns out to be functorial, i.e. module
morphisms extend to morphisms of the twisted modules. As an instance of this fact let us
consider \textit{twisted covariant derivatives} \cite{AschieriCastellani2009}
on $\mathcal{X}_\star$. Those are
left $\mathcal{X}_\star$-linear maps
$\nabla^\mathcal{F}\colon\Xi_\star\otimes_{\mathbb{K}[[\nu]]}\mathcal{T}_\star
\rightarrow\mathcal{T}_\star$ which are compatible with the $\otimes_{\mathcal{X}_\star}$
tensor product structure in the sense that
\begin{equation}\label{eq03}
    \nabla^\mathcal{F}_X(T_1\otimes_{\mathcal{X}_\star} T_2)
    =[\overline{\mathcal{R}}_1\rhd\nabla^\mathcal{F}_{\overline{\mathcal{R}}_{2'}\rhd X}(\overline{\mathcal{R}_{2''}}\rhd T_1)]\otimes_{\mathcal{X}_\star}
    [(\overline{\mathcal{R}}_2\overline{\mathcal{R}}_{1'}\overline{\mathcal{R}}_{1''})
    \rhd T_2]
    +(\overline{\mathcal{R}}_1\rhd T_1)\otimes_{\mathcal{X}_\star}\nabla^\mathcal{F}_{
    \overline{\mathcal{R}}_2\rhd X}T_2
\end{equation}
for all $X\in\Xi_\star$ and $T_1,T_2\in\mathcal{T}_\star$. We further require that
$\nabla^\mathcal{F}_Xf=\mathcal{L}^\mathcal{F}_Xf$ and
\begin{equation}\label{eq04}
    \nabla^\mathcal{F}_X\langle Y,\omega\rangle_\star
    =\langle\overline{\mathcal{R}}_1\rhd\nabla^\mathcal{F}_{\overline{\mathcal{R}}_{2'}\rhd X}(\overline{\mathcal{R}_{2''}}\rhd Y),
    (\overline{\mathcal{R}}_2\overline{\mathcal{R}}_{1'}\overline{\mathcal{R}}_{1''})
    \rhd\omega\rangle_\star
    +\langle\overline{\mathcal{R}}_1\rhd Y,\nabla^\mathcal{F}_{
    \overline{\mathcal{R}}_2\rhd X}\omega\rangle_\star
\end{equation}
for all $X,Y\in\Xi_\star$, $\omega\in\Omega_\star$ and $f\in\mathcal{X}_\star$,
meaning that $\nabla^\mathcal{F}$ should respect the underlying twisted Cartan calculus.
We define \textit{torsion} and \textit{curvature} of a twisted covariant derivative as
the left $\mathcal{X}_\star$-linear maps
$\mathrm{T}^\mathcal{F}_\star\colon\Xi_\star\otimes_{\mathcal{X}_\star}\Xi_\star
\rightarrow\Xi_\star$ and
$\mathrm{R}^\mathcal{F}_\star\colon\Xi_\star\otimes_{\mathcal{X}_\star}\Xi_\star
\otimes_{\mathcal{X}_\star}\Xi_\star
\rightarrow\Xi_\star$
such that
\begin{equation}
\begin{split}
    \mathrm{T}^\mathcal{F}_\star(X,Y)
    :=&\nabla^\mathcal{F}_XY
    -\nabla^\mathcal{F}_{\overline{\mathcal{R}}_1\rhd Y}
    (\overline{\mathcal{R}}_2\rhd X)-[X,Y]_\star,\\
    \mathrm{R}^\mathcal{F}_\star(X,Y,Z)
    :=&\nabla^\mathcal{F}_X\nabla^\mathcal{F}_YZ
    -\nabla^\mathcal{F}_{\overline{\mathcal{R}}_1\rhd Y}
    \nabla^\mathcal{F}_{\overline{\mathcal{R}}_2\rhd X}Z
    -\nabla^\mathcal{F}_{[X,Y]_\star}Z
\end{split}
\end{equation}
for all $X,Y,Z\in\Xi_\star$. One proves that
$\mathrm{T}^\mathcal{F}_\star(X,Y)=-\mathrm{T}^\mathcal{F}_\star(\overline{\mathcal{R}}_1\rhd Y,\overline{\mathcal{R}}_2\rhd X)$ and
$\mathrm{R}^\mathcal{F}_\star(X,Y,Z)=-\mathrm{R}^\mathcal{F}_\star(\overline{\mathcal{R}}_1\rhd Y,\overline{\mathcal{R}}_2\rhd X,Z)$. In other words, torsion and curvature are
completely determined by elements
$\mathrm{T}^\mathcal{F}\in\Omega^2_\star\otimes_{\mathcal{X}_\star}\Xi_\star$
and 
$\mathrm{R}^\mathcal{F}\in\Omega_\star\otimes_{\mathcal{X}_\star}\Omega^2_\star
\otimes_{\mathcal{X}_\star}\Xi_\star$ with
\begin{equation}
    \mathcal{T}^\mathcal{F}_\star(X,Y)=\langle
    X\otimes_{\mathcal{X}_\star}Y,\mathrm{T}^\mathcal{F}\rangle_\star,~
    \mathrm{R}^\mathcal{F}_\star(X,Y,Z)=\langle
    X\otimes_{\mathcal{X}_\star}Y\otimes_{\mathcal{X}_\star}Z,
    \mathrm{R}^\mathcal{F}\rangle_\star
\end{equation}
for all $X,Y,Z\in\Xi_\star$. A \textit{metric} is a left $\mathcal{X}_\star$-linear
non-degenerate map
$\mathbf{g}_\star\colon\Xi_\star\otimes_{\mathcal{X}_\star}\Xi_\star
\rightarrow\mathcal{X}_\star$ such that $\mathbf{g}_\star(Y,X)=
\mathbf{g}_\star(\overline{\mathcal{R}}_1\rhd X,\overline{\mathcal{R}}_2\rhd Y)$
for all $X,Y\in\Xi_\star$. Each metric $\mathbf{g}_\star$
induces a braided-symmetric tensor
$\mathbf{g}=\mathbf{g}^a\otimes_\mathcal{X}\mathbf{g}_a
=\mathbf{g}^A\otimes_{\mathcal{X}_\star}\mathbf{g}_A
\in\Omega_\star\otimes_{\mathcal{X}_\star}\Omega_\star$ by
\begin{equation} \label{Defgstar}
    \mathbf{g}_\star(X,Y)=\langle X\star\langle Y,
    \mathbf{g}^A\rangle_\star,\mathbf{g}_A\rangle_\star.
\end{equation}
A twisted covariant derivative $\nabla^\mathcal{F}$ is said to be
\textit{Levi-Civita} with respect to a metric $\mathbf{g}_\star$ if 
$\nabla^\mathcal{F}\mathbf{g}=0$ and $\mathrm{T}^\mathcal{F}_\star=0$.
As in the classical setting we define the 
\textit{Ricci tensor} as the contraction
$\mathrm{Ric}^\mathcal{F}_\star(X,Y):=\langle \theta^i,\mathrm{R}^\mathcal{F}_\star(e_i,X,Y)\rangle'_\star$, where
$\langle\omega,X\rangle'_\star:=\langle\overline{\mathcal{R}}_1\rhd X,
\overline{\mathcal{R}}_2\rhd\omega\rangle_\star$ for all $X,Y\in\Xi_\star$ and
$\omega\in\Omega_\star$. Note that $\mathrm{Ric}^\mathcal{F}_\star$ is independent of
the choice of dual $\star$-frames $\{e_i\}$, $\{\theta^i\}$.

We recall from \cite{FioreWeber} how to twist deform a classical covariant derivative
$\nabla\colon\Xi\otimes_\mathbb{K}\Xi\rightarrow\Xi$ into a twisted covariant derivative.
First consider the following Lie subalgebra
\begin{equation}
    \mathfrak{e}:=\{Z\in\Xi~|~\mathcal{L}_Z\nabla_XY=\nabla_{[Z,X]}Y+\nabla_X[Z,Y]
    \text{ for all }X,Y\in\Xi\}
\end{equation}
of $\Xi$, called the \textit{equivariance Lie algebra} of $\nabla$. It follows that
$\xi\rhd\nabla_XY=\nabla_{\xi_{(1)}\rhd X}(\xi_{(2)}\rhd Y)$ for all
$\xi\in U\mathfrak{e}$ and $X,Y\in\Xi$. Consider a twist $\mathcal{F}$ on
$U\mathfrak{e}$. Then, according to \cite{FioreWeber}~Proposition~2, the twist deformation
\begin{equation}\label{eq05}
    \nabla^\mathcal{F}_XY:=\nabla_{\overline{\mathcal{F}}_1\rhd X}
    (\overline{\mathcal{F}}_2\rhd Y),~X,Y\in\Xi_\star
\end{equation}
extends to a twisted covariant derivative
$\nabla^\mathcal{F}\colon\Xi_\star\otimes_{\mathbb{K}[[\nu]]}\mathcal{T}_\star
\rightarrow\mathcal{T}_\star$ on $\mathcal{X}_\star$. Moreover, $\nabla^\mathcal{F}$
is $U\mathfrak{e}^\mathcal{F}$-equivariant, i.e. $\xi\rhd\nabla^\mathcal{F}_XT
=\nabla^\mathcal{F}_{\xi_{\widehat{(1)}}\rhd X}(\xi_{\widehat{(2)}}\rhd T)$ and
the compatibility conditions (\ref{eq03}) and (\ref{eq04}) simplify to the expressions
\begin{equation}
\begin{split}
    \nabla^\mathcal{F}_X(T_1\otimes_{\mathcal{X}_\star}T_2)
    =&\nabla^\mathcal{F}_XT_1\otimes_{\mathcal{X}_\star}T_2
    +(\overline{\mathcal{R}}_1\rhd T_1)\otimes_{\mathcal{X}_\star}\nabla^\mathcal{F}_{
    \overline{\mathcal{R}}_2\rhd X}T_2,\\
    \nabla^\mathcal{F}_X\langle Y,\omega\rangle_\star
    =&\langle\nabla^\mathcal{F}_XY,\omega\rangle_\star
    +\langle\overline{\mathcal{R}}_1\rhd Y,
    \nabla^\mathcal{F}_{\overline{\mathcal{R}}_2\rhd X}\omega\rangle_\star
\end{split}
\end{equation}
for all $\xi\in U\mathfrak{e}$, $X,Y\in\Xi_\star$, $T,T_1,T_2\in\mathcal{T}_\star$ and
$\omega\in\Omega_\star$.  For a classical (pseudo-)Riemannian manifold $(M,\mathbf{g})$
with Levi-Civita covariant derivative $\nabla\colon\Xi\otimes_\mathbb{K}\Xi\rightarrow\Xi$
a further specification is obtained via the Lie algebra
of \textit{Killing vector fields} $\mathfrak{k}\subseteq\mathfrak{e}\subseteq\Xi$,
defined by
\begin{equation}
    \mathfrak{k}:=\{Z\in\Xi~|~\mathcal{L}_Z\mathbf{g}(X,Y)
    =\mathbf{g}([Z,X],Y)+\mathbf{g}(X,[Z,Y])
    \text{ for all }X,Y\in\Xi\}.
\end{equation}
In Proposition~3 of \cite{FioreWeber} it is proven that for a twist $\mathcal{F}$ 
on $U\mathfrak{k}$  the  $\mathcal{X}_\star$-bilinear metric 
(\ref{Defgstar}) reduces to  $\mathbf{g}_\star(X,Y)=
\mathbf{g}(\overline{\mathcal{F}}_1\rhd X,\overline{\mathcal{F}}_2\rhd Y)$, 
 and the twist deformation (\ref{eq05}) of the Levi-Civita connection
$\nabla$ of $(M,\mathbf{g})$ is the unique twisted Levi-Civita covariant derivative for
the $\mathbf{g}_\star$. The curvature $\mathrm{R}^\mathcal{F}$ of $\nabla^\mathcal{F}$
is undeformed. Summarizing, in order to provide twist deformations of (Levi-Civita)
covariant derivatives we have to determine Drinfel'd twists based on the Lie algebra
of (Killing) equivariant vector fields.

\section{Twist deformation of smooth submanifolds of \texorpdfstring{$\mathbb{R}^n$}{Rn}}\label{Chap3}

In this section we examine twisted differential geometry on a codimension $k$ submanifold $M$ of  the type (\ref{DefIdeal}). 
Actually, the same constructions with the same twist hold for each submanifold  $M_c:=f_c^{-1}(\{0\})$  in the
$k$-parameter family introduced there.
We write $\mathcal{X}:=\mathcal{C}^\infty(\mathcal{D}_f)$ and
\begin{equation}
    \mathcal{X}^{M_c}:=\mathcal{X}/\mathcal{C}^c
    =\{[g]:=g+\mathcal{C}^c~|~g\in\mathcal{X}\},
\end{equation}
where $\mathcal{C}^c\subseteq\mathcal{X}$ denotes the ideal of smooth functions 
vanishing on $M_c$. It is proven in \cite{FioreWeber}~Theorem~1 that $\mathcal{C}^c
=\bigoplus_{a=1}^k\mathcal{X}\cdot f^a_c=\bigoplus_{a=1}^kf^a_c\cdot\mathcal{X}$, i.e. $\mathcal{C}^c$ is spanned by the components of $f_c$. A similar result (Theorem~1 in \cite{FioFraWebquadrics}) holds in the setting of algebraic submanifolds of  
$\mathbb{R}^n$, i.e. if the $f^a(x)$ are polynomial functions and we define  $\mathcal{X}$ as the algebra
of polynomial functions on $\mathbb{R}^n$.
The Lie algebra
of vector fields on $\mathcal{D}_f$ is denoted by
$\Xi:=\{X^i\partial_i~|~X^i\in\mathcal{X}\}$, where we abbreviate $\partial_i
=\frac{\partial}{\partial x^i}$. There are two Lie subalgebras and 
$\mathcal{X}$-sub-bimodules $\Xi_{\mathcal{C}\mathcal{C}^c}\subseteq\Xi_{\mathcal{C}^c}
\subseteq\Xi$, defined by
\begin{equation}
    \Xi_{\mathcal{C}^c}:=\{X\in\Xi~|~X(\mathcal{C}^c)\subseteq\mathcal{C}^c\}~~~~~
    \text{ and }~~~~~
    \Xi_{\mathcal{C}\mathcal{C}^c}:=\{X\in\Xi~|~X(\mathcal{X})\subseteq\mathcal{C}^c\},
\end{equation}
respectively. Furthermore $\Xi_{\mathcal{C}\mathcal{C}^c}=\bigoplus_{a=1}^kf^a_c\cdot\Xi$
is a Lie ideal in $\Xi_{\mathcal{C}^c}$ and thus the quotient Lie algebra
\begin{equation}
    \Xi^{M_c}:=\Xi_{\mathcal{C}^c}/\Xi_{\mathcal{C}\mathcal{C}^c}
    :=\{[X]:=X+\Xi_{\mathcal{C}\mathcal{C}^c}~|~X\in\Xi_{\mathcal{C}^c}\}
\end{equation}
is an $\mathcal{X}^{M_c}$-bimodule, identified with the vector fields on $M_c$.
In case $c=0$ we suppress the index and simply write $\mathcal{X}^M$, $\Xi^M$,
et cetera. We further define
\begin{equation}
    \Xi_t:=\{X\in\Xi~|~X(f^a)=0\text{ for all }a=1,\ldots,k\}
\end{equation}
the Lie subalgebra and $\mathcal{X}$-sub-bimodule of vector fields that are tangent not only to $M$, but to \textit{all} submanifolds $M_c$
(in fact $X(f^a_c)=0$ for all $X\in\Xi_t$ and $c\in f(\mathcal{D}_f)$). By Proposition 6 in
\cite{FioreWeber}, each element $[X]\in\Xi^{M_c}$ contains a representative $X_t\in\Xi_t$, the tangential projection of $X$.
The requirement that the   algebras in {\it both} vertical columns of (\ref{Diag}) 
are isomorphic as $\mathbb{K}[[\nu]]$-modules - i.e. the basic requirement of deformation quantization applied to both the
algebra of functions on $\RR^n$ and that on $M$ - and the commutativity of the diagram (\ref{Diag}) can be satisfied if the Drinfel'd twist $\mathcal{F}$ is based on $U\Xi_t$, i.e if $\Xi_t$ replaces $\Xi$ in (\ref{twist}); as a bonus, the same holds for all other $M_c$.
In fact, then $\alpha\star f^a=\alpha f^a=f^a\star \alpha$
for all $\alpha\in\mathcal{X}$ and $a=1,..,k$, implying that  also
 $\mathcal{C}_\star$, $\mathcal{C}[[\nu]]$ are isomorphic as $\mathbb{K}[[\nu]]$-modules\footnote{In fact, then
all $\gamma\equiv \sum_{a=1}^kf^a \gamma^a\in \mathcal{C}$ ($\gamma^a\in\mathcal{X}$) can be written also in the form
$\gamma=\sum_{a=1}^kf^a \star \gamma^a$, so that for all $\alpha\in\mathcal{X}$, by the associativity of $\star$, 
$ \gamma\star \alpha=(\sum_{a=1}^kf^a \star \gamma^a)\star \alpha=\sum_{a=1}^kf^a \star (\gamma^a\star \alpha)
=\sum_{a=1}^kf^a (\gamma^a\star \alpha)\in \mathcal{C}[[\nu]]$, as claimed; and similarly for $\alpha\star \gamma$.}.
(On the contrary, using a twist based on $U\Xi^{M}$ would only lead to  \ $\alpha\star f^a\!-\!\alpha f^a\in \mathcal{C}$, $f^a\star \alpha\!-\!f^a\alpha\in \mathcal{C}$, which is  not sufficient  to obtain the same results.)
Adopting  a  twist
$\mathcal{F}$ based on $U\Xi_t$, we obtain deformations of all previously defined spaces.
Namely, $\Xi_{\mathcal{C}\mathcal{C}^c\star}\subseteq\Xi_{\mathcal{C}^c\star}
\subseteq\Xi_\star$ and $\Xi_{t,\star}
$ are $\star$-Lie algebras and $U\Xi_t^\mathcal{F}$-equivariant
$\mathcal{X}_\star$-bimodules, while $\Xi^{M_c}_\star$ is a $\star$-Lie algebra and an 
$U\Xi_t^\mathcal{F}$-equivariant $\mathcal{X}^{M_c}_\star$-bimodule. 
There is an isomorphism $\Xi^{M_c}_\star
\cong\Xi_{\mathcal{C}^c\star}/\Xi_{\mathcal{C}\mathcal{C}^c\star}$ of
$\mathbb{K}[[\nu]]$-modules, i.e. deforming commutes with taking the submanifold quotient
(c.f. \cite{FioreWeber}~Proposition~9). 
As described in the previous section, we obtain the
$U\Xi_t^\mathcal{F}$-equivariant $\mathcal{X}_\star$-bimodule $\Omega_\star$
and the
$U\Xi_t^\mathcal{F}$-equivariant $\mathcal{X}^{M_c}_\star$-bimodule
$\Omega_{M_c\star}$, $\star$-dual to $\Xi_\star$ and $\Xi_{M_c\star}$, respectively.
Moreover, the $\star$-orthogonal module corresponding to tangent vector fields is
the $U\Xi_t^\mathcal{F}$-equivariant $\mathcal{X}_\star$-sub-bimodule
$\Omega_{\perp\star}\subseteq\Omega_\star$ defined by
\begin{equation}\label{eq06}
    \Omega_{\perp\star}:=\{\omega\in\Omega_\star~|~
    \langle\Xi_{t\star},\omega\rangle_\star=0\},
\end{equation}
which is characterized by $\Omega_{\perp\star}=\bigoplus_{a=1}^k\mathcal{X}_\star\star\mathrm{d}f^a
=\bigoplus_{a=1}^k\mathrm{d}f^a\star\mathcal{X}_\star$.

Given a (pseudo-)Riemannian metric $\mathbf{g}=\mathbf{g}^\alpha\otimes\mathbf{g}_\alpha
\in\Omega\otimes_\mathcal{X}\Omega$ on $\mathcal{D}_f$ (by definition  $\mathbf{g}$ is non-degenerate and flip-symmetric) we can further consider
the $\mathbf{g}$-orthogonal spaces
\begin{equation}\label{eq07}
    \Xi_\perp:=\{X\in\Xi~|~\mathbf{g}(X,\Xi_t)=0\}\quad
    \text{ and }\quad
    \Omega_t:=\{\omega\in\Omega~|~\mathbf{g}^{-1}(\omega,\Omega_\perp)=0\},
\end{equation}
where $\mathbf{g}^{-1}=\mathbf{g}^{-1\alpha}\otimes\mathbf{g}^{-1}_\alpha
\in\Xi\otimes_\mathcal{X}\Xi$ is the \textit{inverse metric} and 
$\Omega_\perp$ denotes the classical limit of (\ref{eq06}).
There is a maximal open subset $\mathcal{D}'_f\subseteq\mathcal{D}_f$ such that
$\mathbf{g}_\perp^{-1}:=\mathbf{g}^{-1}\colon
\Omega_\perp\otimes_\mathcal{X}\Omega_\perp\rightarrow\mathcal{X}$ is non-degenerate.
Note that $\mathcal{D}'_f=\mathcal{D}_f$ if $\mathbf{g}$ is Riemannian. If 
in the following $\mathcal{D}'_f\neq\mathcal{D}_f$ we restrict all involved submanifolds
to $M_c\subseteq\mathcal{D}'_f$, so we can assume $\mathcal{D}'_f=\mathcal{D}_f$.
For a twist $\mathcal{F}$ on $U\Xi_t$ the deformations of (\ref{eq07}) read
\begin{equation}
    \Xi_{\perp\star}:=\{X\in\Xi_\star~|~\mathbf{g}_\star(X,\Xi_{t\star})=0\}\quad
    \text{ and }\quad
    \Omega_{t,\star}:=\{\omega\in\Omega_\star~|~\mathbf{g}_\star^{-1}
    (\omega,\Omega_{\perp\star})=0\}.
\end{equation}
According to \cite{FioreWeber}~Proposition~10 we obtain a convenient direct sum 
decomposition in case $\mathcal{F}$ is a twist based on Killing vector fields
$U\mathfrak{k}$: as $\mathfrak{X}_\star$-bimodules
\begin{equation}\label{eq08}
    \Xi_\star\cong\Xi_{t\star}\oplus\Xi_{\perp\star}\quad
    \text{ and }\quad
    \Omega_\star\cong\Omega_{t\star}\oplus\Omega_{\perp\star}
\end{equation}
with $\langle \Xi_{{\scriptscriptstyle{\perp}}\star},\Omega_{t\star}\rangle_\star\!=\!\{0\}$, 
$\Xi_{t\star},\Omega_{{\scriptscriptstyle{\perp}}\star},\Xi_{{\scriptscriptstyle{\perp}}\star},\Omega_{t\star}$
 coincide  with  $\Xi_t[[\nu]],\Omega_{\perp}[[\nu]],\Xi_{\perp}[[\nu]],\Omega_t[[\nu]]$ \ as $\mathbb{K}[[\nu]]$-modules. Similarly for $\star$-tensor (and $\star$-wedge) powers.
The projections $\mathrm{pr}_{t\star}\colon\Xi_\star\rightarrow\Xi_{t\star}$,
$\mathrm{pr}_{\perp\star}\colon\Xi_\star\rightarrow\Xi_{\perp\star}$,
$\mathrm{pr}_{t\star}\colon\Omega_\star\rightarrow\Omega_{t\star}$,
$\mathrm{pr}_{\perp\star}\colon\Omega_\star\rightarrow\Omega_{\perp\star}$,
are $U\mathfrak{k}^\mathcal{F}$-equivariant maps that $\mathbb{K}[[\nu]]$-linearly extend their classical limits 
$\mathrm{pr}_t,\mathrm{pr}_\perp$; they are uniquely extended to $\star$-tensor (and $\star$-wedge) powers.
Furthermore, $\Omega_{t\star}=\{\omega\in\Omega_\star~|~
\langle\Xi_{\perp\star},\omega\rangle_\star=0\}$, and the restrictions
$\mathbf{g}_{t\star},\mathbf{g}_{\perp\star},\mathbf{g}^{-1}_{t\star},
\mathbf{g}^{-1}_{\perp\star}$ of the metric and its inverse to tangent and normal
vector fields, respectively $1$-form, are non-degenerate.
As a consequence, the \textit{first fundamental form}
\begin{equation}
    \mathbf{g}_t^\mathcal{F}:=(\mathrm{pr}_{t\star}\otimes_{\mathcal{X}_\star}
    \mathrm{pr}_{t\star})(\mathbf{g})
    =(\mathrm{pr}_{t}\otimes_{\mathcal{X}}
    \mathrm{pr}_{t})(\mathbf{g})
    =\mathbf{g}_t
\end{equation}
is undeformed.

We continue to describe the dual picture, namely twisted differential $1$-forms on the
submanifolds $M_c$. There we think of tangent vector fields as vector fields on
$M_c$, so with regard to the direct sum decomposition (\ref{eq08})
the following is natural. Setting $\Omega_{\mathcal{C}^c\star}
:=\{\omega\in\Omega_\star~|~\langle\Xi_{\perp\star},\omega\rangle
\subseteq\mathcal{C}^c[[\nu]]\}$ and $\Omega_{\mathcal{C}\mathcal{C}^c\star}
:=\bigoplus_{a=1}^k\Omega_\star\star f^a_c=\bigoplus_{a=1}^kf^a_c\star\Omega_\star$ we obtain
\begin{equation}
    \Omega_{M_c\star}
    =\Omega_{\mathcal{C}^c\star}/\Omega_{\mathcal{C}\mathcal{C}^c\star}
    =\{[\omega]=\omega+\Omega_{\mathcal{C}\mathcal{C}^c\star}~|~
    \omega\in\Omega_{\mathcal{C}^c\star}\}.
\end{equation}
It turns out, c.f. \cite{FioreWeber}~Proposition~11, that for every
$X\in\Xi_{\mathcal{C}^c\star}$ and $\omega\in\Omega_{\mathcal{C}^c\star}$ we have
\begin{equation}
    \mathrm{pr}_{t\star}(X)\in[X]\in\Xi_{M_c\star}
    \text{ and }
    \mathrm{pr}_{t\star}(\omega)\in[\omega]\in\Omega_{M_c\star}.
\end{equation}
In other words, for every $[X]\in\Xi_{M_c\star}$ and $[\omega]\in\Omega_{M_c\star}$
we can find representatives $\mathrm{pr}_{t\star}(X)$ and 
$\mathrm{pr}_{t\star}(\omega)$ in $\Xi_{t\star}$ and $\Omega_{t\star}$,
respectively.

Consider the Levi-Civita connection $\nabla$ on $(\mathcal{D}_f,\mathbf{g})$.
In the following we describe the twisted Riemannian geometry on the $k$-parameter
family $M_c$ of codimension $k$ smooth submanifolds.
We already mentioned that for a twist $\mathcal{F}$ on $U\mathfrak{k}$ the twist
deformation $\nabla^\mathcal{F}$ is the twisted Levi-Civita connection with respect to
$\mathbf{g}_\star$. This induces a \textit{twisted second fundamental form}
\begin{equation}
    \Pi_\star^\mathcal{F}:=\mathrm{pr}_{\perp\star}\circ\nabla^\mathcal{F}|_{
    \Xi_{t\star}\otimes_{\mathcal{X}_\star}\Xi_{t\star}}
    \colon\Xi_{t\star}\otimes_{\mathcal{X}_\star}\Xi_{t\star}\rightarrow\Xi_{\perp\star}
\end{equation}
and twisted Levi-Civita connection
\begin{equation}
    \nabla_t^\mathcal{F}:=\mathrm{pr}_{t\star}\circ\nabla^\mathcal{F}|_{
    \Xi_{t\star}\otimes_{\mathbb{K}[[\nu]]}\Xi_{t\star}}
    \colon\Xi_{t\star}\otimes_{\mathbb{K}[[\nu]]}\Xi_{t\star}\rightarrow\Xi_{t\star}
\end{equation}
on $M_c$. It is proven in \cite{FioreWeber}~Proposition~12 that the tensors
corresponding to the second fundamental form, curvature, Ricci tensor and Ricci
scalar of $\nabla^\mathcal{F}_t$ remain undeformed, i.e.
\begin{equation}
\begin{split}
    \Pi^\mathcal{F}=&
    \Pi\in(\Omega_t\otimes_\mathcal{X}\Omega_t\otimes_\mathcal{X}\Xi_\perp)[[\nu]],\\
    \mathrm{Ric}_t^\mathcal{F}
    =&\mathrm{Ric}_t\in(\Omega\otimes_\mathcal{X}\Omega)[[\nu]],
\end{split}
\hspace{1cm}
\begin{split}
    \mathrm{R}_t^\mathcal{F}=&\mathrm{R}_t
    \in(\Omega_t\otimes_\mathcal{X}\Omega_t^2\otimes_\mathcal{X}\Xi_t)[[\nu]],\\
    \mathfrak{R}^\mathcal{F}=&\mathfrak{R}\in\mathcal{X}.
\end{split}
\end{equation}
However, the corresponding linear maps combine via the
\textit{twisted Gauss equation}
\begin{equation}
\begin{split}
    \mathbf{g}_\star(\mathrm{R}_\star^\mathcal{F}(X,Y,Z),W)
    =&\mathbf{g}_\star(\mathrm{R}_{t\star}^\mathcal{F}(X,Y,Z),W)
    +\mathbf{g}_\star(\Pi_\star^\mathcal{F}(X,\overline{\mathcal{R}}_1\rhd Z),
    \Pi_\star^\mathcal{F}(\overline{\mathcal{R}}_2\rhd Y,W))\\
    &-\mathbf{g}_\star(
    \Pi_\star^\mathcal{F}(\overline{\mathcal{R}}_{1\widehat{(1)}}\rhd Y,
    \overline{\mathcal{R}}_{1\widehat{(2)}}\rhd Z),
    \Pi_\star^\mathcal{F}(\overline{\mathcal{R}}_2\rhd X,W))
\end{split}
\end{equation}
for all $X,Y,Z,W\in\Xi_{t\star}$, c.f. \cite{FioreWeber}~Proposition~13.

\subsection{Twisted differential calculus on algebraic submanifolds by generators and relations}\label{TwistedDC}

In this section we describe the twisted differential calculus on algebraic submanifolds 
$M_c$ in terms of generators and relations. We choose the convenient description
via the \textit{differential calculus algebra}, which allows us to describe
functions, differential forms, vector fields and their interaction simultaneously.
The construction is divided into two parts, where we first describe the calculus
algebra on $\mathbb{R}^n$ and afterwards quotient by an ideal to achieve the
description of the submanifolds. Denote the Cartesian coordinate functions of
$\mathbb{R}^n$ by $x^i$ and further abbreviate $\xi^i=\mathrm{d}x^i$,
$\partial_i=\frac{\partial}{\partial x^i}$. The unit function is denoted by
$x^0=\mathbf{1}$ and $\eta^i\in\{x^i,\xi^i,\partial_i\}$ can denote the $i$-th
coordinate function, $1$-form or coordinate vector field.
Those are the generators of our constructions and consequently we focus on the
subalgebra $\mathcal{X}:=\mathrm{Pol}^\bullet(\mathbb{R}^n)
\subseteq\mathcal{C}^\infty(\mathbb{R}^n)$ of polynomial functions and vector fields
$\Xi:=\{h^i\partial_i~|~h^i\in\mathrm{Pol}^\bullet(\mathbb{R}^n)\}$ with polynomial
coefficients in this section.
Here and in the following Latin indices $i,j,k,\ldots$ run over $1,\ldots,n$,
while Greek indices $\mu,\nu,\rho,\ldots$ run over $0,1,\ldots,n$.
The \textit{differential calculus algebra of $\mathbb{R}^n$}  
is the associative unital $*$-algebra $\mathcal{Q}^\bullet$ generated by
the Hermitian elements $\{x^0,x^i,\xi^i,\mathrm{i}\partial_i\}$ modulo the
relations
\begin{equation}\label{eq09}
\begin{split}
    &x^0\eta^i-\eta^i=\eta^ix^0-\eta^i=0\\
    &x^ix^j-x^jx^i=0\\
    &\xi^ix^j-x^j\xi^i=0\\
    &\partial_i\partial_j-\partial_j\partial_i=0\\
    &\partial_i\xi^j-\xi^j\partial_i=0\\
    &\xi^i\xi^j+\xi^j\xi^i=0\\
    &\partial_ix^j-x^j\partial_i-\delta_i^jx^0=0.
\end{split}
\end{equation}
For any Lie subalgebra $\mathfrak{g}\subseteq\mathrm{aff}(n)$ (the affine Lie algebra on  $\mathbb{R}^n$)  we obtain a left
$U\mathfrak{g}$-module algebra action on $\mathcal{Q}^\bullet$, determined
by primitive elements $g\in\mathfrak{g}$ on generators by
\begin{equation}
    g\rhd x^0=\epsilon(g)x^0,~
    g\rhd x^i=x^\mu\tau^{\mu i}(g),~
    g\rhd\xi^i=\xi^j\tau^{ji}(g),~
    g\rhd\partial_i=\tau^{ij}(S(g))\partial_j.
\end{equation}
The action is well-defined since $\mathrm{aff}(n)$ preserves the ideal (\ref{eq09}).
A basis of $\mathcal{Q}^\bullet$ is
\begin{equation}
    \mathcal{B}:=\{\beta^{\vec{p},\vec{q},\vec{r}}
    :=(\xi^1)^{p_1}\ldots(\xi^n)^{p_n}(x^1)^{q_1}\ldots(x^n)^{q_n}
    \partial_1^{r_1}\ldots\partial_n^{r_n}~|~
    \vec{p}\in\{1,0\}^n,\vec{q},\vec{r}\in\mathbb{N}_0^n\}.
\end{equation}
Introducing the total degrees
$p:=\sum_{i=1}^np_i$, $q:=\sum_{i=1}^nq_i$ and $r:=\sum_{i=1}^nr_i$ we can define 
gradings $\natural,\sharp$ on $\mathcal{Q}^\bullet$ compatible with the $*$-algebra structure of the latter
by setting $\natural(\beta^{\vec{p},\vec{q},\vec{r}}):=p$, $\sharp(\beta^{\vec{p},\vec{q},\vec{r}}):=q-r$
on the elements of $\mathcal{B}$.
There are three fundamental subalgebras
$\mathcal{X}=\bigoplus_{q=0}^\infty\mathcal{X}^q$, 
$\Lambda^\bullet
=\bigoplus_{p=0}^n\Lambda^p$, 
$\mathcal{D}=\bigoplus_{r=0}^\infty\mathcal{D}^r$ 
of $\mathcal{Q}^\bullet$, where $\mathcal{X}^q,\Lambda^p,\mathcal{D}^r$ denote
the homogeneous polynomials in, respectively, $x^i$, $\xi^i$ and $\partial_i$
and we set $\mathcal{X}^0=\Lambda^0=\mathcal{D}^0=\mathbb{C}\cdot x^0$.
Then $\mathcal{X}=\biguplus_{q=0}^\infty\tilde{\mathcal{X}}^q$ and
$\mathcal{D}=\biguplus_{r=0}^\infty\tilde{\mathcal{D}}^r$
are filtered with respect to the inhomogeneous polynomials 
$\tilde{\mathcal{X}}^q:=\bigoplus_{h=0}^q\mathcal{X}^h$
and $\tilde{\mathcal{D}}^r:=\bigoplus_{h=0}^r\mathcal{D}^h$, respectively.
We further define the left 
$U\mathfrak{g}$-$*$-modules $\mathcal{Q}^{pqr}:=\Lambda^p\tilde{\mathcal{X}}^q
\tilde{\mathcal{D}}^r$ with basis
$\mathcal{B}^{pqr}:=\{\beta^{\vec{p},\vec{q},\vec{r}}~|~p=\sum_{i=1}^np_i,~
\sum_{i=1}^nq_i\leq q,~\sum_{i=1}^nr_i\leq r\}$.
Then $\mathcal{Q}^\bullet$ is $p$-graded and filtered by $q$ and $r$ with decomposition
\begin{equation}
    \mathcal{Q}^\bullet=\bigoplus_{p=0}^\infty\biguplus_{q=0}^\infty
    \biguplus_{r=0}^\infty\mathcal{Q}^{pqr}.
\end{equation}
For a real or unitary twist $\mathcal{F}$ on $U\mathfrak{g}$ it turns out that the
left $U\mathfrak{g}_\mathcal{F}$-module $*$-algebra $\mathcal{Q}^\bullet_\star$
is again described in terms of generators and relations, namely
\begin{equation}\label{eq12}
\begin{split}
    &x^0\star x^i-x^i=x^i\star x^0-x^i=0\\
    &x^0\star \xi^i-\xi^i=\xi^i\star x^0-\xi^i=0\\
    &x^0\star\partial'_i-\partial'_i=\partial'_i\star x^0-\partial'_i=0\\
    &x^i\star x^j-x^\nu\star x^\mu R_{\mu\nu}{}^{ij}=0\\
    &\xi^i\star x^j-x^\nu\star\xi^hR_{h\nu}{}^{ij}=0\\
    &\partial'_i\star\partial'_j-R_{\mu\nu}{}^{hk}\partial'_k\star\partial'_h=0\\
    &\partial'_i\star\xi^j-\xi^j\star\partial'_i=0\\
    &\xi^i\star\xi^j+\xi^k\star\xi^hR_{hk}{}^{ij}=0\\
    &\partial'_i\star x^j-R_{\mu i}{}^{jk}x^\mu\star\partial'_k-\delta_i^jx^0=0,
\end{split}
\end{equation}
where $\partial'_i:=S(\beta)\rhd\partial_i=\tau^{ij}(\beta)\partial_j$
is the $\star$-dual frame to $\xi^i=\mathrm{d}x^i$, transforming via
$g\rhd\partial'_i=\tau^{ij}(S_\mathcal{F}(g))\partial'_j$.
We further denoted 
$R_{\mu\nu}{}^{ij}:=(\tau^{\mu i}\otimes\tau^{\nu j})(\mathcal{R})$.
Eq. (\ref{eq12}) are the analogue of the  relations
 defining the quantum group equivariant `quantum spaces' introduced in \cite{FRT} and the associated differential calculi algebras (see e.g. formulae (1.10-15)
in \cite{Fio04JPA}). 
As in the untwisted case, $\mathcal{Q}^\bullet_\star$ is $p$-graded and
filtered by $q$ and $r$ with decomposition
\begin{equation}
    \mathcal{Q}^\bullet_\star=\bigoplus_{p=0}^\infty\biguplus_{q=0}^\infty
    \biguplus_{r=0}^\infty\mathcal{Q}^{pqr}_\star,
\end{equation}
where $\mathcal{Q}^{pqr}_\star:=\Lambda^p_\star\tilde{\mathcal{X}}^q_\star
\tilde{\mathcal{D}}^r_\star$ consists of (in)homogeneous $\star$-polynomials
with basis $\mathcal{B}^{p,q,r}_\star$. The $*$-involution on 
$\mathcal{Q}^\bullet_\star$ is undeformed if $\mathcal{F}$ is real and
in case $\mathcal{F}$ is unitary it is defined on generators by 
\begin{equation}
    (x^0)^{*_\star}:=x^0,~~~~~
    (x^i)^{*_\star}:=x^\mu\tau^{\mu i}(S(\beta)),~~~~~
    (\xi^i)^{*_\star}:=\xi^j\tau^{ji}(S(\beta)),~~~~~
    (\partial'_i)^{*_\star}:=-\tau^{ij}(\beta^{-1})\partial'_j,
\end{equation}
which follow from the general formula $s^{*_\star}    =S(\beta)\rhd s^*$.
Now we induce a twist quantization of the submanifolds $M_c$
corresponding to the common zero sets $f^a_c(x)=f^a(x)-c^a=0$ for all $a=1,\ldots,k$.
Choose a basis $\{e_1,\ldots,e_B\}$ of $\mathfrak{g}$ and the corresponding
structure constants $C^\gamma_{\alpha\beta}\in\mathcal{X}$. Instead of
$\{\partial_1,\ldots,\partial_n\}$ we can consider $\{e_1,\ldots,e_B,e_{B+1},\ldots,
e_{B+k}\}$
with $e_{B+a}:=\sum_{i=1}^n\frac{\partial f^a}{\partial x^i}\partial_i$
as a complete set of vector fields $\Xi$ with relations
\begin{equation}\label{eq10}
\begin{split}
    &e_{B+a}x^h-x^he_{B+a}-\frac{\partial f^a}{\partial x^h}=0,\qquad   a=1,\ldots,k\\
    &e_\alpha x^h-x^he_\alpha-x^\mu\tau^{\mu h}(e_\alpha)=0,\qquad  \alpha=1,\ldots,B\\
    &t^\alpha_\ell e_\alpha=0,\qquad  \ell=1,\ldots,B+k-n\\
    &e_\alpha e_\beta-e_\beta e_\alpha-C^\gamma_{\alpha\beta}e_\gamma=0,  \\
    &e_\alpha\xi^i-\xi^ie_\alpha=0 
\end{split}
\end{equation}
for some $t_\ell^\alpha\in\mathcal{X}$.
Consider the free algebra $\mathcal{A}^{'\bullet}$ generated by
$x^0,\ldots,x^n,\xi^1,\ldots,\xi^n,e_1,\ldots,e_B$. Similarly to the previous
discussion one shows that
$\mathcal{A}^{'\bullet}=\bigoplus_{p=0}^\infty\biguplus_{q=0}^\infty
\biguplus_{r=0}^\infty\mathcal{A}^{'pqr}$ is a $p$-graded, $q,r$-filtered
algebra. We denote the ideal in $\mathcal{A}^{'\bullet}$ generated by
the usual relations on $x^i,\xi^i$, the relations (\ref{eq10}) for
$\alpha\leq B$ and $f^a_c(x)=0=\mathrm{d}f^a$
by $\mathcal{I}_{M_c}$. The corresponding differential calculus algebra is
$\mathcal{Q}^\bullet_{M_c}:=\mathcal{A}^{'\bullet}/\mathcal{I}_{M_c}$.
It is graded and filtered according to
\begin{equation}
    \mathcal{Q}^\bullet_{M_c}=
    \bigoplus_{p=0}^{n-1}\biguplus_{q=0}^\infty
    \biguplus_{r=0}^\infty\mathcal{Q}^{pqr}_{M_c}
\end{equation}
where $\mathcal{Q}^{pqr}_{M_c}:=\mathcal{A}^{'pqr}/\mathcal{I}^{pqr}_{M_c}$
and $\mathcal{I}^{pqr}_{M_c}:=\mathcal{I}_{M_c}\cap\mathcal{A}^{'pqr}$.
One shows that $\mathcal{Q}^\bullet_{M_c}$ is a left $U\mathfrak{g}$-module
$*$-algebra.
For a real or unitary twist $\mathcal{F}$ on $U\mathfrak{g}$ the
\textit{twisted differential calculus algebra} $\mathcal{Q}^\bullet_{M_c\star}$
on $M_c$ can be defined as the result of either path of the commuting diagram
\begin{equation}\label{eq13}
\begin{tikzcd}
(\mathcal{A}^{'\bullet},\mathcal{I}_{M_c}) \arrow{d}[swap]{\mathcal{F}}
\arrow{rr}{\text{quotient}}
& & \mathcal{Q}^\bullet_{M_c} \arrow{d}{\mathcal{F}}\\
(\mathcal{A}^{'\bullet}_\star,\mathcal{I}_{M_c\star})
\arrow{rr}{\text{quotient}}
& & \mathcal{Q}^\bullet_{M_c\star}
\end{tikzcd}
\end{equation}
i.e. twist deformation and the quotient procedure commute. 
It is $p$-graded and $q,r$-filtered via the left 
$U\mathfrak{g}_\mathcal{F}$-$*$-submodules $\mathcal{Q}^{pqr}_{M_c\star}$, i.e.
\begin{equation}
    \mathcal{Q}^\bullet_{M_c\star}=
    \bigoplus_{p=0}^{n-1}\biguplus_{q=0}^\infty
    \biguplus_{r=0}^\infty\mathcal{Q}^{pqr}_{M_c\star}.
\end{equation}
The generators and relations determining $\mathcal{Q}^\bullet_{M_c\star}$
are precisely the twist deformations of the generators and relations of
$\mathcal{Q}^\bullet_{M_c}$.

\subsection{Twisted quadrics in \texorpdfstring{$\mathbb{R}^3$}{R3}}\label{TwistedQuad}

The determining function of quadric surfaces of $\mathbb{R}^3$ is
$f(x)=\frac{1}{2}a_{ij}x^ix^j+a_{0i}x^i+\frac{1}{2}a_{00}$
with $a_{\mu\nu}=a_{\nu\mu}$ for $\mu,\nu=0,1,2,3$.
Defining $f_i:=\frac{\partial f}{\partial x^i}=a_{ij}x^j+a_{0i}$ and
$L_{ij}:=f_i\partial_j-f_j\partial_i$ gives a complete set
$S_L:=\{L_{ij}\}_{i,j=1,\ldots,n}$ of tangent vector fields.
Since
\begin{equation}
    [L_{ij},L_{hk}]
    =a_{jh}L_{ik}-a_{ih}L_{jk}-a_{jk}L_{ih}+a_{ik}L_{jh},
\end{equation}
$S_L$ is a Lie algebra $\mathfrak{g}$, which is acting on $\mathcal{X}$ via
\begin{equation}
    L_{ij}\rhd x^h
    =(a_{ik}x^k+a_{0i})\delta^h_j
    -(a_{jk}x^k+a_{0j})\delta^h_i,
\end{equation}
i.e. $\mathfrak{g}\subseteq\mathrm{aff}(n)$ is a Lie subalgebra of the affine Lie
algebra. Following the procedure of Section~\ref{TwistedDC},
starting from the differential calculus algebra $\mathcal{Q}^\bullet$ of $\mathbb{R}^n$
with relations (\ref{eq12}) we first obtain the differential calculus algebra
$\mathcal{Q}^\bullet_M$ on the quadric surface $M$ with relations (\ref{eq13}).
By an Euclidean coordinate transformation 
we can make $a_{ij}=a_i\delta_{ij}$, 
$a_{0i}=0$ if  $a_i\neq 0$ (quadrics in canonical form).

Given a twist $\mathcal{F}$ on $U\mathfrak{g}$ we then get a quantization
$\mathcal{Q}^\bullet_{M\star}$ of the quadric surface. The latter is deformed as a
$*$-algebra if $\mathcal{F}$ is unitary or real.
In \cite{FioFraWebquadrics} this is exemplified via Abelian and Jordanian twist
deformations of all quadric surfaces of $\mathbb{R}^3$, except the ellipsoid.
The results are summarized in Figure~\ref{QuadricSummary}.
\begin{figure}[h!]
\begin{center}
\begin{tabular}{|c|c|c|c|c|c|c|c|c|c|c|}
\hline
& $a_1$ & $a_2$ & $a_3$ & $a_{03}$ & $a_{00}$ & $r$ &  quadric  &$\mathfrak{g}\simeq$ & Abelian & Jordanian \\
\hline
(a) & $+$ & 0 & 0 & $-$ & & 3 &  parabolic cylinder & $\mathfrak{h}(1)$ & Yes & No \\
\hline
(b) & $+$ & $+$ & 0  & $-$ & & 4 &
elliptic paraboloid & $\mathfrak{so}(2)\ltimes\mathbb{R}^2$ & Yes & No \\
\hline
(c) & $+$ & $+$ & 0  & 0 &  $-$ & 3 & elliptic cylinder & 
\begin{tabular}{c}
$\mathfrak{so}(2) \cross \mathbb{R}^2$\\ 
$\mathfrak{so}(2) \times \mathbb{R}$ 
\end{tabular}
& \begin{tabular}{c}
Yes\\ 
Yes
\end{tabular}
 &  \begin{tabular}{c}
No\\ 
No
\end{tabular} \\
\hline
(d) & $+$ & $-$ & 0 & $-$ &  & 4 & hyperbolic paraboloid  & $\mathfrak{so}(1,\!1)\!\ltimes\!\mathbb{R}^2 $ 
& Yes & Yes \\
\hline
(e) & $+$ & $-$ & 0 & 0 & $-$  & 3 & hyperbolic cylinder  & 
\begin{tabular}{c}
$\mathfrak{so}(1,\!1) \cross \mathbb{R}^2$\\ 
$\mathfrak{so}(1,\!1) \!\times\! \mathbb{R}$ 
\end{tabular}
& \begin{tabular}{c}
Yes\\ 
Yes
\end{tabular}
 &  \begin{tabular}{c}
Yes\\ 
No
\end{tabular} \\
\hline
(f) & $+$ & $+$ & $-$  & 0 & $-$ & 4 &  1-sheet  hyperboloid
& $\mathfrak{so}(2,1)$ & No & Yes \\
\hline
(g) & $+$ & $+$ & $-$  & 0 &+& 4 & 2-sheet hyperboloid
& $\mathfrak{so}(2,1)$ & No & Yes \\
\hline
(h) & $+$ & $+$ & $-$  & 0 & 0 & 3 &  elliptic cone$^\dagger$
& $\mathfrak{so}(2,\!1)\!\times\!\mathbb{R}$ & Yes$^\dagger$ & Yes \\
\hline
(i) & $+$ & $+$ & $+$ & 0 & $-$ & 4 & ellipsoid & $\mathfrak{so}(3)$ & No & No \\
\hline
\end{tabular}
\end{center}
\caption{Overview of the quadrics in $\mathbb{R}^3$: signs of the coefficients of the equations in canonical form
(if not specified, all $a_{00}\in\mathbb{R}$ are possible), 
rank,  associated symmetry Lie algebra $\mathfrak{g}$, type of twist deformation;
$\mathfrak{h}(1)$  stands for the Heisenberg algebra.
For fixed $a_i$ each class gives a family of submanifolds $M_c$ parametrized by $c$,
except classes (f), (g), (h), which altogether give a single family; 
so there are 7 families of submanifolds. We can always make $a_1=1$
by a rescaling of $f$. The $\dagger$ reminds  that the
cone (e) is not a single closed manifold, due to the singularity in the apex.}
\label{QuadricSummary}
\end{figure}

\subsubsection*{Twisted differential geometry on the hyperboloids and cone}

Let us recall the family of hyperboloids in Minkowski $\mathbb{R}^3$ in detail.
For positive numbers $a,b>0$ and $c\in\mathbb{R}$ we consider the solutions
$x\in\mathbb{R}^3$ of the equation
\begin{equation}
    f_c(x)=\frac{1}{2}((x^1)^2+a(x^2)^2-b(x^3)^2)-c=0
\end{equation}
and denote their collection by $M_c$.
$M_{c>0}$ is a family of $1$-sheet hyperboloids and
$M_{c<0}$ a family of $2$-sheet hyperboloids. Together they from a foliation
$\{M_c\}_{c\in\mathbb{R}\setminus\{0\}}$ of $\mathbb{R}^3\setminus M_0$,
where $M_0$ constitutes the cone\footnote{By removing the origin we consider $M_0$ as a smooth submanifold of
$\mathbb{R}^3$ consisting of two disconnected components.}.
The submanifolds $M_c$ have an $\mathfrak{g}=\mathfrak{so}(2,1)$ symmetry with base
vectors $L_{12}=x^1\partial_2-ax^2\partial_1$,
$L_{13}=x^2\partial_3+bx^3\partial_1$ and
$L_{23}=ax^2\partial_3+bx^3\partial_2$. In fact,
$H:=\frac{2}{\sqrt{b}}L_{13}$ and
$E^\pm:=\frac{1}{\sqrt{a}}L_{12}\pm\frac{1}{\sqrt{ab}}L_{23}$ satisfy
\begin{equation}
    [H,E^\pm]=\pm2E^\pm,~~~~~[E^+,E^-]=-H.
\end{equation}
For computational reasons it is convenient to work in the coordinate system given by
the eigenvectors $y^\pm:=x^1\pm\sqrt{b}x^3$ and $y^0:=x^2$
of $H$ corresponding to the eigenvalues $\lambda^\pm=\pm 2$ and $\lambda^0=0$.
The associated coordinate $1$-forms and vector fields are
$\eta^\pm=\mathrm{d}y^\pm$, $\eta^0=\mathrm{d}y^0$ and $\partial_\pm=
\frac{\partial}{\partial y^\pm}$, $\partial_0=\frac{\partial}{\partial y^0}$.
In this coordinate system we have
\begin{equation}\label{eq15}
\begin{split}
    f_c(y)=\frac{1}{2}y^+y^-+\frac{a}{2}(y^0)^2-c,\\
    H=2y^+\partial_+-2y^-\partial_-~,~~~~~
    E^\pm=\frac{1}{\sqrt{a}}y^\pm\partial_0-2\sqrt{a}y^0\partial_\mp.
\end{split}
\end{equation}
For later use we also define $\partial^\pm=2a\partial_\mp$ and $\partial^0=\partial_0$.
With this choice of basis the $U\mathfrak{g}$-action on 
$u^i\in\{y^i,\partial^i,\eta^i\}$, $i=+,-,0$, is determined by
\begin{equation}
    H\rhd u^i=\lambda u^i,~~~~~~~~~
    E^\pm\rhd u^i=\delta_0^i\frac{1}{\sqrt{a}}u^\pm-\delta^i_\mp\sqrt{a}u^0.
\end{equation}
We consider the unitary twist 
$\mathcal{F}=\exp(H/2\otimes\mathrm{log}(1+\mathrm{i}\nu E^+))
\in U\mathfrak{g}^{\otimes 2}[[\nu]]$ and its deformed Hopf algebra
$U\mathfrak{g}_\mathcal{F}$. The latter coincides with the $\mathbb{C}[[\nu]]$-linear
extension of the algebra $U\mathfrak{g}$, with $\mathbb{C}[[\nu]]$-linear extended
counit but twisted coproduct $\Delta_\mathcal{F}$ and antipode $S_\mathcal{F}$
determined by
\begin{align*}
    \Delta_\mathcal{F}(H)=\Delta(H)-\mathrm{i}\nu H\otimes\frac{E^+}{1+\mathrm{i}\nu E^+},
    ~~~~~\Delta_\mathcal{F}(E^+)=\Delta(E^+)+\mathrm{i}\nu E^+\otimes E^+,\\
    \Delta_\mathcal{F}(E^-)=\Delta(E^-)
    -\frac{\mathrm{i}\nu}{2}H\otimes\bigg(H+\frac{\mathrm{i}\nu E^+}{1+\mathrm{i}\nu E^+}\bigg)\frac{1}{1+\mathrm{i}\nu E^+}
    -\mathrm{i}\nu E^-\otimes\frac{E^+}{1+\mathrm{i}\nu E^+}
    -\frac{\nu^2}{4}H^2\otimes\frac{E^+}{(1+\mathrm{i}\nu E^+)^2},\\
    S_\mathcal{F}(H)=S(H)(1+\mathrm{i}\nu E^+),
    ~~~~~S_\mathcal{F}(E^+)=\frac{S(E^+)}{1+\mathrm{i}\nu E^+},\\
    S_\mathcal{F}(E^-)=S(E^-)(1+\mathrm{i}\nu E^+)
    -\frac{\mathrm{i}\nu}{2}H(1+\mathrm{i}\nu E^+)\bigg(H+\frac{\mathrm{i}\nu E^+}{1+\mathrm{i}\nu E^+}\bigg)
    +\frac{\nu^2}{4}H(1+\mathrm{i}\nu E^+)HE^+.
\end{align*}
The corresponding twist deformation $\mathcal{Q}^\bullet_\star$ of the differential
calculus algebra of $\mathbb{R}^3$ is the free algebra $\star$-generated by
$u^i\in\{y^i,\partial^i,\xi^i\}$, $i=+,-,0$,
with $\star$-product of $u^i\in\{y^i,\partial^i,\xi^i\}$ and
$w^j\in\{y^j,\partial^j,\xi^j\}$, $i,j=+,-,0$, given by
\begin{equation}
    u^i\star w^j=u^iw^j
    +\mathrm{i}\nu(\delta^i_--\delta^i_+)u^i
    \bigg(\frac{1}{\sqrt{a}}\delta^j_0w^+-2\sqrt{a}\delta^j_-w^0\bigg)
    +\delta^i_+\delta^j_-2\nu^2u^+w^+.
\end{equation}
One can formulate the differential calculus algebra only in terms of these generators and the relations
\begin{align*}
 u^+\!\star\! u^0=u^0\!\star\! u^+\!-\!\frac{\mathrm{i}\nu}{\sqrt{a}}u^+\!\star\! u^+, 
\qquad  u^+\!\star\! u^-=u^-\!\star\! u^+\!+\!2 \mathrm{i}\nu \sqrt{a}\, u^0\!\star\! u^+
\!+\! 2\nu^2u^+\!\star\! u^+, \\[8pt]
 u^0\!\star\! u^-=u^-\!\star\! u^0\!-\!\frac{\mathrm{i}\nu}{\sqrt{a}}u^-\!\star\! u^+, 
\qquad u^+\!\star\! \eta^+=\eta^+\!\star\! u^+, \quad 
u^+\!\star\! \eta^0=\eta^0\!\star\! u^+-\frac{\mathrm{i}\nu}{\sqrt{a}}\eta^+\!\star\! u^+,\\[8pt]
u^+\!\star\! \eta^-=\eta^-\!\star\! u^++2 \mathrm{i}\nu \sqrt{a}\, \eta^0\!\star\! u^+\!+\! 2\nu^2\eta^+\!\star\! u^+,
\qquad  u^0\!\star\! \eta^+=\eta^+\!\star\! u^0+\frac{\mathrm{i}\nu}{\sqrt{a}}\eta^+\!\star\! u^+ ,  \\[8pt]
u^0\!\star\! \eta^0=\eta^0\!\star\! u^0, \quad 
u^0\!\star\! \eta^-=\eta^-\!\star\! u^0-\frac{\mathrm{i}\nu}{\sqrt{a}}\eta^-\!\star\! u^+,\quad
u^-\!\star\! \eta^+=\eta^+\!\star\! u^--2\mathrm{i}\nu \sqrt{a}\,\eta^+\!\star\! u^0,\\[8pt] 
u^-\!\star\! \eta^0=\eta^0\!\star\! u^-+\frac{\mathrm{i}\nu}{\sqrt{a}}\eta^+\!\star\! u^-
+2\nu^2\eta^+\!\star\! u^0,\\[8pt] 
u^-\!\star\! \eta^-=\eta^-\!\star\! u^-+2\mathrm{i}\nu \sqrt{a}\big(\eta^-\!\star\! u^0-\eta^0\!\star\! u^-\big)+2\nu^2 \,\eta^-\!\star\! u^+
\end{align*}
for $u^i=y^i,\partial^i$, $i=+,-,0$. 
The twisted Leibniz rule for the derivatives read
\begin{align*}
\partial^+\!\star\! y^+=y^+\!\star\! \partial^+, \quad  \partial^0\!\star\! y^+
=y^+\!\star\! \partial^0+\frac{\mathrm{i}\nu}{\sqrt{a}}y^+\!\star\! \partial^+ , \quad
\partial^-\!\star\! y^+=2a+y^+\!\star\! \partial^--\mathrm{i}2\nu \sqrt{a}y^+\!\star\! \partial^0,\\[8pt] 
\partial^+\!\star\! y^0=y^0\!\star\! \partial^+-\frac{\mathrm{i}\nu}{\sqrt{a}}y^+\!\star\! \partial^+,\qquad 
\partial^-\!\star\! y^0=y^0\!\star\! \partial^-+\mathrm{i}2\nu\sqrt{a}+
\frac{\mathrm{i}\nu}{\sqrt{a}}y^+\!\star\! \partial^-
+2\nu^2y^+\!\star\! \partial^0,\\[8pt] 
 \partial^0\!\star\! y^0=1+y^0\!\star\! \partial^0, \qquad \qquad\quad \:
\partial^+\!\star\! y^-=2a+y^-\!\star\! \partial^++\mathrm{i}2 \nu \sqrt{a}\, y^0\!\star\! \partial^+
+2 \nu^2 y^+\!\star\! \partial^+,\\[8pt]
\partial^0\!\star\! y^-=y^-\!\star\! \partial^0-\frac{\mathrm{i}\nu}{\sqrt{a}}y^-\!\star\! \partial^+, 
\qquad
\partial^-\!\star\! y^-=y^-\!\star\! \partial^-+\mathrm{i}2 \nu \sqrt{a}\big(y^-\!\star\! \partial^0-y^0\!\star\! \partial^-\big)+2\nu^2 \,y^-\!\star\! \partial^+,
\end{align*}
while  the twisted wedge products fulfill
\begin{align*}
\eta^+\!\star\! \eta^+=0, \qquad 
&\eta^0\!\star\! \eta^0=0,\qquad 
&\eta^-\!\star\! \eta^-=2\mathrm{i}\nu \sqrt{a}\, \eta^0\!\star\! \eta^-,\\[8pt]
\eta^+\!\star\! \eta^0+\eta^0\!\star\! \eta^+=0,\quad 
&\eta^+\!\star\! \eta^-+\eta^-\!\star\! \eta^+=2\mathrm{i}\nu\sqrt{a}\,\eta^+\!\star\! \eta^0 ,\quad &
\eta^0\!\star\! \eta^-+\eta^-\!\star\! \eta^0=\frac{\mathrm{i}\nu}{\sqrt{a}}\eta^-\!\star\! \eta^+.
\end{align*}
In terms of star products
\begin{align*}
    H    =2(\partial_+\star y^+-1
    -y^-\star\partial_-),~~~~~~
    E^\pm    = \displaystyle\frac{1}{\sqrt{a}}\partial_0\star y^\pm
    -2\sqrt{a}y^0\star\partial_\mp.
\end{align*}
The relations  characterizing the $U\mathfrak{g}^\mathcal{F}$-equivariant 
$*$-algebra  $\mathcal{Q}_{M_c\star}^\bullet$ become
\begin{align*}
    0 =&f_c(y)\equiv
    \frac{1}{2}y^-\star y^++\frac{a}{2}y^0\star y^0-c,\\[6pt]
  0=& \mathrm{d}f_c =\frac{1}{2}(y^-\star\eta^++\eta^-\star y^+)    +ay^0\star\eta^0,\\[6pt]
  0= &y^-\star E^+-y^+\star E^--\sqrt{a}\,y^0\star H+\mathrm{i}\nu y^+\star H-2\mathrm{i}\nu (1+\mathrm{i}\nu) y^+\star E^+.
\end{align*}
The $*$-structures on $U\mathfrak{g}^\mathcal{F}$, $\mathcal{Q}^\bullet_\star, \mathcal{Q}_{M_c\star}^\bullet$  remain undeformed except
$(u^-)^{*_\star}
=(u^-)^*-2\mathrm{i}\nu\sqrt{a}(u^0)^*$.

\subsubsection*{Twisted Riemannian geometry on the circular hyperboloids}

Let us consider the Minkowski metric
$\mathbf{g}:=\mathrm{d}x^1\otimes\mathrm{d}x^1+\mathrm{d}x^2\otimes\mathrm{d}x^2
-\mathrm{d}x^3\otimes\mathrm{d}x^3$ on $\mathbb{R}^3$ with corresponding constants
$\mathbf{g}(\partial_i,\partial_j)=\eta_{ij}$. This metric is invariant under
$\mathfrak{so}(2,1)$ and so it is equivariant under the induced
$U\mathfrak{so}(2,1)$-action.
In this last section we specify the previous discussion on hyperboloids to the family
$M_c=f^{-1}_c(\{0\})$ of circular hyperboloids and cone, where
\begin{equation}
    f_c(x)=\frac{1}{2}((x^1)^2+(x^2)^2-(x^3)^2)-c,
\end{equation}
or equivalently (\ref{eq15}) with $a=1$ in the transformed coordinates.
The Lie $*$-algebra symmetry $\mathfrak{g}\cong\mathfrak{so}(2,1)$ of $M_c$ is spanned
by $L_{12},L_{13},L_{23}$, or equivalently $H,E^+,E^-$. Depending on the sign of $c$, the first fundamental form
$\mathbf{g}_t=\mathbf{g}\circ(\mathrm{pr}_t\otimes\mathrm{pr}_t)$ structures
$M_c$ as a Riemannian (for $c<0$) or a Lorentzian (for $c>0$) manifold.
On the cone $M_0$ there is a degeneracy of $\mathbf{g}$. Furthermore,
for $c\neq 0$ the second
fundamental form is $\Pi(X,Y)=-\frac{1}{2c}\mathbf{g}(X,Y)V_\perp$ for $X,Y\in\Xi_t$,
where $V_\perp=(\partial_jf_c)\eta^{ji}\partial_i=x^i\partial_i$.
Choosing a basis $v_1,v_2$ of $\Xi_t$ and setting
$\mathbf{g}_{\alpha\beta}=\mathbf{g}(v_\alpha,v_\beta)$ the Gauss theorem determines
the curvature, Ricci tensor and Ricci scalar on $M_c$ by
\begin{equation}\label{eq14}
    \mathrm{R}_t{}^\delta_{\alpha\beta\gamma}
    =\frac{\mathbf{g}_{\alpha\gamma}\delta^\delta_\beta
    -\mathbf{g}_{\beta\gamma}\delta^\delta_\alpha}{2c},~~~~~
    \mathrm{Ric}_t{}_{\beta\gamma}
    =\mathrm{R}_t{}^\alpha_{\alpha\beta\gamma}
    =-\frac{\mathbf{g}_{\beta\gamma}}{2c},~~~~~
    \mathfrak{R}_t=\mathrm{Ric}_t{}_{\beta\beta}=-\frac{1}{c}.
\end{equation}
This implies that $M_{c<0}$ is a de Sitter space $dS_2$ and
$M_{c>0}$ consists of two copies of anti-de Sitter spaces $AdS_2$. In the limit
$c\rightarrow 0$ the expressions (\ref{eq14}) diverge.
Now $\{H,E^\pm\}$ is a complete set of vector fields on $M_c$ with linear dependence
relation $y^-E^+-y^+E^--y^0H=0$, where we employed again the coordinate system
$y^\pm:=x^1\pm\sqrt{b}x^3$ and $y^0:=x^2$ of eigenvectors of $H$.
As before we consider the twisted differential calculus algebra
$\mathcal{Q}^\bullet_{M_c\star}$ for the unitary Jordanian twist
$\mathcal{F}=\exp(H/2\otimes\mathrm{log}(1+\mathrm{i}\nu E^+))$.
Following Section~\ref{sectRG} the tensors (\ref{eq14}) remain undeformed under the
twist, while
\begin{align*}
\Pi^\mathcal{F}_\star(X,Y)=-\frac {1}{2c}\,\mathbf{g}_{t\star}(X,Y)\,V_\perp=-\frac {1}{2c}\,\mathbf{g}_{t\star}(X,Y)\star V_\perp
\end{align*}
holds using the $U\mathfrak{k}$-invariance of $V_\perp$. 
Similarly
\begin{align*}
\mathrm{R}^\mathcal{F}_{t\star}(X,Y,Z)=
\frac{(\overline{\mathcal{R}}_1\rhd Y) \star\mathbf{g}_{t\star}(\overline{\mathcal{R}}_2\rhd  X,Z)-X\star \mathbf{g}_{t\star}(Y,Z)}{2c},\qquad \mathrm{Ric}^\mathcal{F}_{t\star}(Y,Z)
=-\frac {\mathbf{g}_{t\star}(Y,Z)}{2c}  
\end{align*}
for all $X,Y,Z\in\Xi_{t\star}$ and we obtain explicit expressions of
$\mathbf{g}_{t\star}$ on the generating vector fields $H,E^\pm$:
\begin{align*}
\mathbf{g}_{t\star}(H,H)=-8y^+y^-, \qquad\mathbf{g}_{t\star}(H,E^\pm)=-2y^\pm y^0,\\[6pt]
\mathbf{g}_{t\star}(E^+,E^+)=(y^+)^2, \qquad\quad  \mathbf{g}_{t\star}(E^+,E^-)=2c+(y^0)^2
-2\mathrm{i}\nu y^+y^0-2\nu^2(y^+)^2, \\[6pt]
\mathbf{g}_{t\star}(E^+,H)=-2y^+y^0+2\mathrm{i}\nu (y^+)^2, \qquad\qquad \mathbf{g}_{t\star}(E^-,E^+)=2c+(y^0)^2,\\[6pt] 
\mathbf{g}_{t\star}(E^-,E^-)=(y^-)^2 ,\quad \mathbf{g}_{t\star}(E^-,H)=-2y^0y^--2\mathrm{i}\nu[2c+(y^0)^2]
+2\mathrm{i}\nu y^0y^-.
\end{align*}
Furthermore, the twisted Levi-Civita connection is determined by
\begin{align*}
   \nabla^\mathcal{F}_{E^+}E^+
    =-2y^+\partial_- ,\qquad
    \nabla^\mathcal{F}_{E^+}E^-= 
    -2y^+\partial_+-2y^0\partial_0
    +4\mathrm{i}\nu\partial_-
    +4\nu^2y^+\partial_- ,\\[6pt] 
\nabla^\mathcal{F}_{E^+}H= 
    4y^0\partial_-    -4\mathrm{i}\nu y^+\partial_-    ,\qquad\qquad\qquad
    \nabla^\mathcal{F}_{E^-}E^+
    =-2y^-\partial_--2y^0\partial_0,\\[6pt]
    \nabla^\mathcal{F}_{E^-}E^-
    =-2y^-\partial_+
    +4\mathrm{i}\nu y^0\partial_+,\qquad
    \nabla^\mathcal{F}_{E^-}H=
    -4y^0\partial_+
    +4\mathrm{i}\nu(y^0\partial_0+y^-\partial_-),\\[6pt]
    \nabla^\mathcal{F}_HE^+=
    2y^+\partial_0,\qquad         
    \nabla^\mathcal{F}_HE^-=
    -2y^-\partial_0,\qquad 
    \nabla^\mathcal{F}_HH
    =4y^+\partial_++4y^-\partial_-
\end{align*}
on the generating vector fields $H,E^\pm$.



\end{document}